\documentclass[useAMS,usenatbib]{mn2e}
\usepackage{color}
\usepackage{colortbl}
\usepackage{dcolumn}
\usepackage{amsmath}
\usepackage{amssymb}
\usepackage{graphicx}
\usepackage{epsfig}
\usepackage{changebar}
\usepackage{natbib}
\usepackage{multirow}
\usepackage{subfigure}
\usepackage{txfonts}
\usepackage{relsize}
\usepackage{verbatim} 
\usepackage{natbib}
\usepackage{rotating}
\usepackage{lscape}
\newcommand {\NH}{NH$_3$}
\newcommand {\water}{H$_2$O}
\newcommand{\HII}{\mbox{HII}}

\newcommand{\ie}{i.e.,}
\newcommand{\lsun}{L$_\odot$}
\newcommand{\msun}{M$_\odot$}

\newcommand{\aap}{A\&A}
\newcommand{\araa}{ARAA}

\newcommand{\mnras}{MNRAS}
\newcommand{\apj}{ApJ}
\newcommand{\apjs}{ApJS}

\newcommand{\pasp}{PASP}
\newcommand{\pasa}{PASA}
\newcommand{\aj}{AJ}

\title[The G305 star-forming complex: UCHII regions]{The G305 star-forming complex: a wide-area radio survey of ultra-compact HII regions}
\author[L.Hindson]{L.Hindson$^{1,2}$\thanks{E-mail:
l.hindson@herts.ac.uk}, M.A. Thompson$^{1}$, J.S.Urquhart$^{2,6}$, A. Faimali$^{1}$, J.S. Clark$^{3}$, B. Davies$^{4,5}$\\
$^{1}$\rm Centre for Astrophysics Research, Science and Technology Research Institute, University of Hertfordshire, College Lane, Hatfield, AL10 9AB, UK\\ \rm $^{2}$ATNF, CSIRO Astronomy and Space Science, P.O. Box 76, Epping, NSW 1710, Australia\\ 
$^{3}$\rm Department of Physics and Astronomy, The Open University, Walton Hall, Milton Keynes, MK7 6AA, UK\\
$^{4}$\rm Institute of Astronomy, University of Cambridge, Madingley Road, Cambridge, CB3 0HA, UK\\
$^{5}$\rm School of Physics \& Astronomy, University of Leeds, Woodhouse Lane, Leeds, LS2 9JT, UK\\
$^{6}$\rm Max-Planck-Institut f\"ur Radioastronomie, Auf dem H\"ugel 69, 53121 Bonn, Germany}

\begin{document}
\date{Accepted. Received ; in original form }

\pagerange{\pageref{firstpage}--\pageref{lastpage}} \pubyear{2010}

\maketitle

\label{firstpage}
\begin{abstract}
We present wide-area radio continuum 5.5 and 8.8\,GHz (5.5 and 3.4\,cm) Australia Telescope Compact Array observations of the complex and rich massive star-forming region G305. The aim of this study is to perform an un-targeted survey of the region in search of the compact radio emission associated with ultra-compact (UC) HII regions. Observations presented here encompass the entire complex and have a maximum resolution of $\sim1.5\times1.4\arcsec$ and sensitivity of $\sim0.07$\,mJy\,beam$^{-1}$. By applying a data reduction method that emphasises small-scale structure, we are able to detect 71 compact radio sources distributed throughout the observed field. To explore the nature of these compact radio sources we compare to mid-infrared data and in this way identify 56 background sources, eight stellar radio sources, a single bright-rimmed cloud and six candidate \mbox{UC\,HII} regions. The physical properties of these candidate \mbox{UC\,HII} regions are determined and reveal five candidates have peak properties consistent with known \mbox{UC\,HII} regions with source radii ranging from 0.04--0.1\,pc, emission measures from 2.56--10.3$\times 10^{-6}$\,pc cm$^{-6}$ and electron densities of 0.34--1.03$\times 10^4$\,cm$^{-3}$. We comment on these sites of recent massive star formation within G305 and by comparing to other star formation tracers (masers, \NH, YSOs) build a picture of the star formation history of the region. Using these results we estimate a lower limit to the star formation rate for the region of $\sim$ 0.003\,M$_{\odot}$\,yr$^{-1}$. .
\end{abstract}


\section{Introduction}

The G305 star-forming complex is one of the most luminous \HII\ regions in the Galaxy, comprising of a number of bright radio sources, with an integrated flux indicative of $>30$ deeply embedded O7 V stars \citep{Clark2004}. The complex is located at \mbox{$l= 305.4\degr$}, \mbox{$b = 0.1\degr$} and a heliocentric distance of $3.8\pm0.6$\,kpc placing it within the Scutum Crux arm of the Milky Way. Star formation appears to commenced with the formation of the two clusters Danks\,1 and 2 approximately 3--6\,Myr ago \citep{Davies2011B}, at the center of the complex, the radiation and stellar winds from massive stars within these clusters appears to have excavated a cavity in the surrounding molecular material. Star formation is ongoing around the periphery of this cavity with numerous HII regions, infrared hotspots, masers and dense molecular structure \citep{Clark2004,Hindson2010}. 

G305 contains several distinct sites and epochs of star formation in close proximity, which allows the investigation of massive star formation, evolution and the impact on the surrounding environment such as triggering \citep{Elmegreen1977, Habing1979, Elmegreen2002}. The morphology of the region and position of star formation indicators suggests that interaction between the evolved massive stars, ongoing star formation and molecular material is taking place. Coupled with its relatively close distance makes G305 an excellent laboratory in which to carry out a multi-wavelength study of massive star formation. Our ultimate aim is to characterise the star formation history of the region and to evaluate the impact of the central cluster and massive stars on the current and future star formation.

A key observable feature in the evolution of a massive star is the radio emission associated with an HII region. These regions of ionised hydrogen form as the young massive star(s), still embedded in the natal molecular cloud, begins to ionise the surroundings via high-energy UV photons. As the HII region evolves and expands it goes through the following observed evolutionary stages: hyper-compact, ultra-compact, compact, classical and diffuse \citep{Churchwell2002}. However, details of the evolutionary sequence and impact on the local environment are still poorly understood. 

The ultra-compact (UC) evolutionary phase of a HII region is defined as a small diameter $(< 0.1$\,pc), high electron density ($> 10^{4} \rm\, cm^{-3}$) and high emission measure ($> 10^7\rm\, pc\, cm^{-6}$) region of the ionised gas that surround the youngest and most massive O and B stars (\citealt{Wood1989A,Wood1989B}; hereafter WC89A and B). With an estimated lifetime of $<10^5$ years \citep{Comeron1996,Davies2011} \mbox{UC\,HII} regions provide a direct tracer of one of the earliest stages of massive star formation. 

Previous radio observations of G305 have either covered small areas at high resolution (e.g.\ around the methanol masers of \citealt{Walsh1998}) or were of low angular resolution and sensitivity \citep{Goss1970,Caswell1987}. Here we report on high resolution and sensitive observations of the 5.5 and 8.8\,GHz radio continuum emission across the G305 complex. We aim to detect and estimate properties of a complete and unbiased sample of \mbox{UC\,HII} regions within the complex. We compare these observations to the clusters Danks\,1 and 2, star formation tracers such as masers, young stellar objects (YSOs) and the dense molecular environment (\citealt{Hindson2010}; hereafter Paper\,1) to improve our understanding of the star formation mechanism(s) and history of the region. 

This is the third in a series of papers aimed at studying massive star formation associated with this region. In two previous publications we have presented observations of the dense molecular material traced by \NH\ emission and star formation traced by \water\ maser emission (Paper\,1) and high resolution near-infrared observations of the evolved massive stars associated with the central clusters Danks\,1 and 2 (\citealt{Davies2011B}; hereafter Paper\,2).

In Section\,2 we present details of the observations and data reduction process, in Section\,3 we present our method for determining the nature of the detected radio emission through comparison with mid-infrared data. In Section\,4 we derive physical properties of the candidate \mbox{UC\,HII} regions to determine if they are characteristic of \mbox{UC\,HII} regions, Section\,5 presents our discussion of massive star formation within G305 and comparison with other star formation tracers and the dense molecular environment and in Section\,6 we present a summary.

\section{Observations \& data reduction procedures}

\begin{table}
\begin{tabular}[h]{ccccc}
 \hline Array & Date & Synthesised & Baseline\\
 		Config&	dd/mm/yy &	Beam (\arcsec) & Range (m) \\
 			
 	\hline
	
 			750B	&	26/02/10	&	$9.3\times10.5 $	& 168 - 4500	\\
 			1.5A	&	28/07/09	&	$4.8\times5.4 $ 	& 153 - 4469	\\
 			H75 	&	10/07/09	&	$87.0\times87.0 $	& 31 - 82(4408)	\\
 	 		6A		&	06/06/09	&	$1.5\times1.4$	 	& 628 - 5939	\\
 			H214 	&	23/05/09	&	$33.6\times33.6 $& 92 - 247(4500)	\\
 			H168 	&	10/05/09	&	$40.8\times40.8 $& 61 - 192(4469)	\\

 \hline
 \end{tabular}
 \caption{Observational parameters for ATCA radio observations. Bracketed baseline ranges for the hybrid arrays represent the sixth antenna.}
 \label{Observing_table}
 \end{table}
 
\subsection{Observations}
Observations of the radio continuum towards G305 were made using the Australia Telescope Compact Array (ATCA), which is located at the Paul Wild Observatory, Narrabri, New South Wales, Australia \footnote{The Australia Telescope Compact Array is part of the Australia Telescope National Facility, which is funded by the Commonwealth of 
Australia for operation as a National Facility managed by CSIRO.}. The ATCA consists of $6\times 22$\,m antennas, five of which lie on a 3\,km east-west (E-W) railway track with the sixth antenna located 3\,km further west. The array also has a 214\,m north spur which allows for compact hybrid array configurations. The antennas can be positioned in several configurations with maximum and minimum baselines of 6\,km and 30\,m respectively. We made use of six array configurations the \mbox{E-W} 6A, 1.5A and 750B long baseline arrays and the H214, H168 and H75 hybrid arrays. The dates of observations and synthesised beams of each array can be found in Table\,1.

The observations were made simultaneously at two IF bands centred at 5.5 and 8.8\,GHz and utilised the wide-band continuum mode of the new Compact Array Broadband Backend (CABB; \citealt{Wilson2011}). This resulted in a bandwidth of 2\,GHz with 2048$\times$1\,MHz channels in each IF. At the longest baseline of 6\,km we obtain a maximum resolution of $1.5\times 1.4 \arcsec$ and at the shortest baseline of 30\,m we are sensitive to emission on a scale of $\sim\, 5$ and $\sim\, 3\arcmin$ at 5.5 and 8.8\,GHz respectively. 

\mbox{E-W} array observations were made over a twelve-hour period and hybrid array configurations over a six-hour period providing good hour angle coverage. To correct for fluctuations in the phase and amplitude, caused by atmospheric and instrumental effects, the total scan time is split into blocks of 15 minutes of on source integration sandwiched between 2-minute observations of the phase calibrator 1352$-$63. For absolute calibration of the flux density the standard flux calibrator 1934$-$638 was observed once during the observation for approximately 10 minutes. To calibrate the bandpass the bright point source 1253$-$055 was also observed during the observations. In order to map the entire G305 region, at Nyquist sampling, the 8.8\,GHz band required a mosaic of 357 individual points in a hexagonal pattern this resulted in over sampling at 5.5\,GHz. With a scan rate of 2s per point (spaced over a range of hour angles in order to improve $uv$ coverage) we could thus complete a map in approximately one hour and observe each pointing centre $\sim\,7$--11 times for each hybrid and \mbox{E-W} configurations respectively giving a total on source integration time on each field of $\sim\,1$ minute. 

\subsection{Data reduction}
\label{sect:dr}
The calibration and reduction of the data were performed using the {\sc MIRIAD} reduction package following standard ATCA continuum procedures. The reduction strategy described below is designed to emphasise the faint and compact radio emission ($<3\arcsec$) associated with \mbox{UC\,HII} regions. We defer the analysis and discussion of the large-scale radio continuum structure of G305 to a following publication.

Our reduction strategy is as follows. Single fields were first Fourier transformed using multi-frequency synthesis mode to improve $uv$ coverage and image fidelity. These individual dirty images were then imaged out to 3 times the primary beam ($\sim\, 5\arcmin$ at 8.8\,GHz and $\sim\,8\arcmin$ at 5.5\,GHz) to account for emission from bright sources outside of the primary beam main lobe. These maps were produced with a pixel size that samples the synthesised beam by three pixels and a robust weighting of 0.5, as opposed to natural or uniform, in order to reduce side lobe levels and produce a uniform synthesised beam shape at the sacrifice of some sensitivity.

An unfortunate consequence of mapping such a large area is the snapshot nature and consequent sparse $uv$ sampling of individual fields. This introduces significant artefacts, such as rippling, side lobes and dark bowls in fields coincident with complex and bright large-scale emission \citep{Taylor1999}. These artefacts can easily be confused with real compact emission and results in the localised noise being up to three times higher towards bright large-scale emission. We are interested in detecting the compact radio emission associated with \mbox{UC\,HII} regions and these artefacts as well as the bright extended emission hamper this goal. Therefore, to emphasise small-scale structure and remove artefacts all baselines $<15$\,k$\lambda$ were cut (cf. \citealt{Kurtz1994,Mucke2002}) from the $uv$ data set which results in the removal of all flux associated with the large-scale emission $\gtrsim\, 0.2\arcmin$. As a result, the integrated flux of resolved sources may be underestimated by as much as 50\% and should be considered as a lower limit. This removal of extended flux does not significantly affect the main goal of this study and is deemed an acceptable consequence of the improved data reduction and compact source detection. 

With the large-scale emission removed, the compact emission dirty maps could be more easily cleaned. A cut off level of $3\times\sigma$, where $\sigma$ is the standard deviation of the noise in each image, was selected resulting in a few hundred to few thousand clean components depending on the emission in the image. Finally the 357 individual restored fields were stitched together using the linear mosaic task {\sc LINMOS}. In this mosaic we obtain a sensitivity of 0.07 and 0.15\,mJy\,beam$^{-1}$ for the 5.5 and 8.8\,GHz maps respectively and a dynamic range of $\sim\,1000$. 

The final compact radio mosaic was analysed in the following way. The peak flux densities and coordinates were determined using the task {\sc MAXFIT} in {\sc MIRIAD}. For unresolved and spherical sources the integrated flux density and size was determined by fitting a Gaussian using the task {\sc IMFIT}. In the case of sources that are resolved and not well fit by a Gaussian, we determine the integrated flux densities and sizes by carefully fitting irregular-shaped apertures around the sources and extracting the flux.

\section{Source Classification}

The radio emission detected at 5.5 and 8.8\,GHz is not only generated by the thermal free-free emission associated with HII regions but may originate from a number of astronomical sources for instance; thermal emission from stellar winds \citep{Kenny2007}, bright-rimmed clouds \citep{Thompson2004} and non-thermal emission from colliding wind binaries \citep{Dough2003}, pulsars \citep{Lazaridis2008}, supernova remnants \citep{Gaensler1997} and active galactic nuclei \citep{Soria2010}. In the following section, we describe how we separate candidate \mbox{UC\,HII} regions from these contaminating sources by using ancillary mid-infrared data.

\begin{figure*}
\includegraphics[width=1.0\textwidth, trim=10 20 40 50]{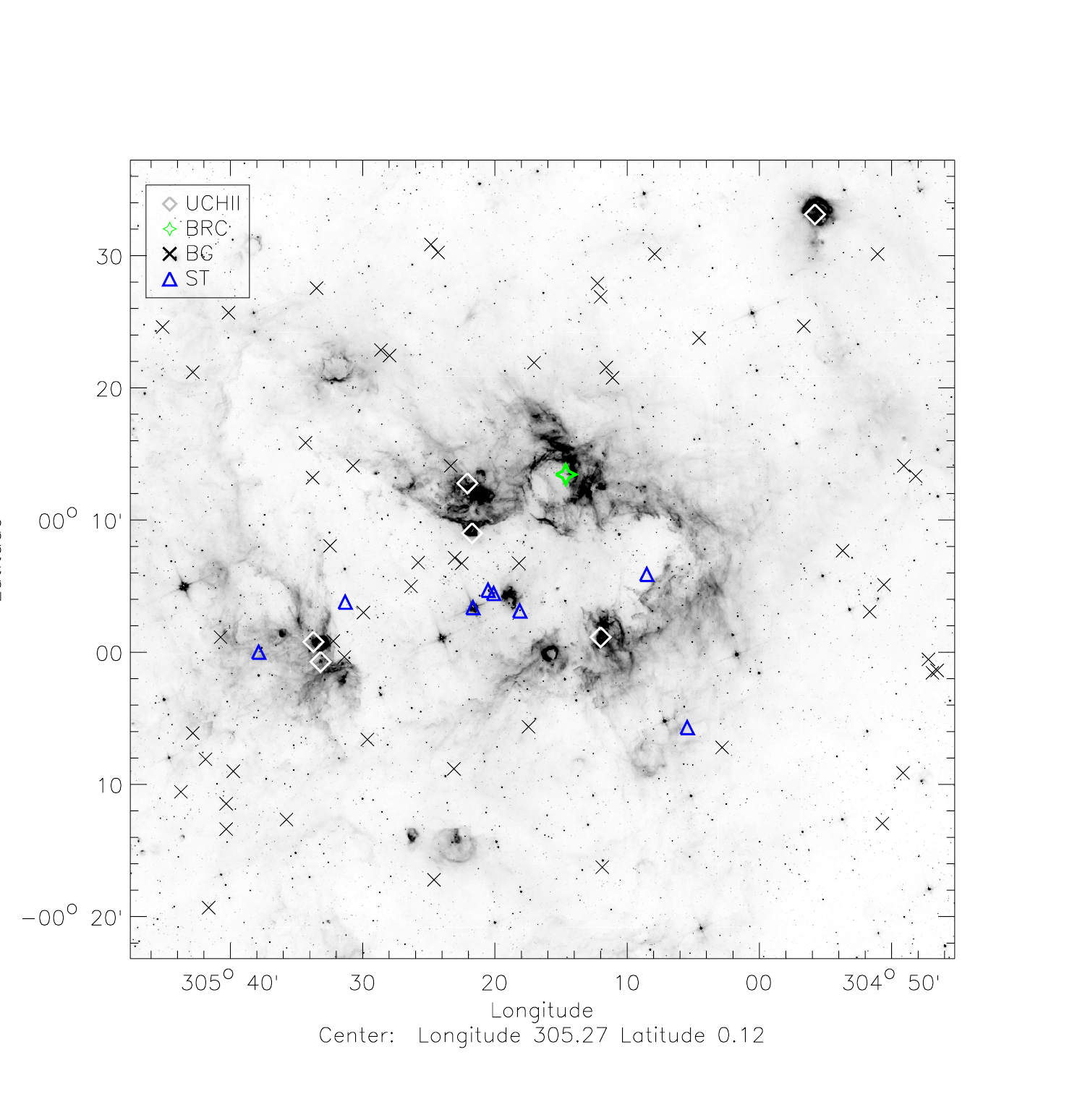} 
\caption{The 71 compact 5.5 GHz radio sources are overlaid onto a 5.8\,$\umu$m GLIMPSE image. White diamonds indicate candidate \mbox{UC\,HII} regions, blue triangles are stellar sources, black crosses show background sources and a green star highlights the bright-rimmed cloud.}
\label{CompactRadioGlimpse}
\end{figure*}

In Fig.\,\ref{CompactRadioGlimpse} we present a mid-infrared image of the G305 region that shows the distribution of the detected compact radio sources. These are randomly distributed across the field, which would suggest many are background sources, rather than being associated with G305. In Table\,2 we present the positions and derived properties of the 71 compact radio sources identified from our radio data. Several sources are not detected in the lower sensitivity 8.8\,GHz maps and for these sources we present the upper limit $3\sigma$ flux level. We find the majority of radio sources detected are unresolved (59) and given that the majority of \mbox{UC\,HII} regions presented in the literature are resolved at similar resolutions (e.g.\ \citealt{Wood1989B, Kurtz1999, Urquhart2007,Urquhart2009}) it is likely that most of these unresolved sources are extragalactic in origin.

\begin{table*}
\centering     
  \begin{tabular}{cccccccccccc}
   \hline 
   Source  & Source	 & \multicolumn{2}{c}{Dimensions}  & \multicolumn{4}{c}{Observed flux density} &Spectral \\ 
	       Index   & Name  & \multicolumn{2}{c}{Observed} & \multicolumn{2}{c}{Peak}  &\multicolumn{2}{c}{Integrated}& Index \\ \cline{3-4} \cline{5-8}
	       	       &          & maj & min &   \multicolumn{2}{c}{(mJy beam$^{-1}$)}       & \multicolumn {2}{c}{(mJy)} &($\alpha$) \\
	        &           & ('')   & ('')    & \textit{f}$_{5.5}$ & \textit{f}$_{8.8}$ & \textit{f}$_{5.5}$ & \textit{f}$_{8.8}$ &    \\
	        \hline
0    &    G305.375+00.112    & \multicolumn{2}{c}{unresolved} &    2.9 $\pm$ 0.3    &    2.2 $\pm$ 0.3    &    3.3    &    2.6    &    $-$0.2	\\
1    &    G305.384+00.119    & \multicolumn{2}{c}{unresolved} &    1.2 $\pm$ 0.3    &    1.5 $\pm$ 0.3    &    1.5    &    1.9    &    0.7 \\
2$^{\rm \textit{st}}$    &    G305.342+00.078    & \multicolumn{2}{c}{unresolved} &    0.9 $\pm$ 0.1    &    1.6 $\pm$ 0.2    &    1.0    &    1.9    &    1.7 \\
3$^{\rm \textit{st}}$    &    G305.361+00.056    &\multicolumn{2}{c}{unresolved}       &    1.8 $\pm$ 0.3    &    2.5 $\pm$ 0.3    &    2.4    &    2.9    &   0.6 \\
4    &    G305.303+00.112    & \multicolumn{2}{c}{unresolved} &    7.4 $\pm$ 0.3    &    6.7 $\pm$ 0.2    &    9.3    &    7.9    &    $-$0.4 \\
5    &    G305.439+00.083    & \multicolumn{2}{c}{unresolved} &    3.1 $\pm$ 0.3    &    4.4 $\pm$ 0.1    &    3.8    &    5.7    &    0.8 \\
6    &    G305.430+00.113    & \multicolumn{2}{c}{unresolved} &    0.9 $\pm$ 0.1    &    1.0 $\pm$ 0.1    &    0.9    &    1.0    &    0.4 \\
7    &    G305.541+00.134    &   \multicolumn{2}{c}{unresolved}    &    0.7 $\pm$ 0.3    &    0.6 $\pm$ 0.1    &    1.0    &    0.6    & $-$1.4 \\
8$^{\rm \textit{uc}}$    &    G305.362+00.150    &    2.40 $\pm$ 0.05    &    3.0 $\pm$ 0.05    &    24.0 $\pm$ 0.8    &    26.0 $\pm$ 0.9    &    29.0    &    34.0    & $-$\\
9$^{\rm \textit{uc}}$ &    G305.368+00.213    &    4.20 $\pm$ 0.09    &    4.2 $\pm$ 0.08    &    28.0 $\pm$ 1.1    &    15.0 $\pm$ 1.0    &    62.0    &    54.0    & $-$\\
10$^{\rm \textit{brc}}$    &    G305.244+00.224    &    3.60 $\pm$ 0.66    &    4.8 $\pm$ 0.36    &    1.7 $\pm$ 0.2    &    0.9 $\pm$ 0.1    &    5.5    &    1.7    & $-$\\
11$^{\rm \textit{st}}$    &    G305.142+00.098    & \multicolumn{2}{c}{unresolved} &    3.5 $\pm$ 0.1    &    3.0 $\pm$ 0.1    &    4.2    &    3.4    &    $-$0.4\\
12    &    G305.193+00.359    & \multicolumn{2}{c}{unresolved} &    2.5 $\pm$ 0.3    &    1.3 $\pm$ 0.1    &    3.1    &    1.8    &    $-$0.6 	\\
13    &    G305.185+00.346    & \multicolumn{2}{c}{unresolved} &    1.2 $\pm$ 0.3    &    0.9 $\pm$ 0.1    &    1.6    &    1.1    &    $-$1.4 	\\
14    &    G305.204+00.465    &    3.60 $\pm$ 1.04    &    6.0 $\pm$ 0.23    &    3.1 $\pm$ 0.4    &    1.5 $\pm$ 0.2    &    8.8    &    2.5    & $-$\\
15    &    G305.563+00.220    & \multicolumn{2}{c}{unresolved} &    2.0 $\pm$ 0.2    &    1.6 $\pm$ 0.2    &    2.3    &    1.8    &    $-$0.2 	\\
16    &    G305.512+00.235    & \multicolumn{2}{c}{unresolved} &    1.7 $\pm$ 0.1    &    1.5 $\pm$ 0.1    &    1.8   &    1.4    &    $-$0.2 	\\
17$^{\rm \textit{uc}}$    &    G305.562+00.013    &    6.00 $\pm$ 1.51    &    5.4 $\pm$ 1.00    &    7.3 $\pm$ 0.8    &    3.4 $\pm$ 0.4    &    31.4    &    12.1    & $-$\\
18    &    G305.537+00.014    &    \multicolumn{2}{c}{unresolved}     &    2.0 $\pm$ 0.1    &    1.7 $\pm$ 0.2    &    2.0    &    1.7    & $-$0.1 \\
19    &    G305.523$-$00.006    & \multicolumn{2}{c}{unresolved} &    4.2 $\pm$ 0.2    &    3.0 $\pm$ 0.2    &    5.3    &    4.7    &    $-$0.2 	\\
20$^{\rm \textit{uc}}$    &    G305.553$-$00.012    &    4.20 $\pm$ 0.18    &    3.6 $\pm$ 0.16    &    8.4 $\pm$ 0.7    &    5.3 $\pm$ 0.4    &    16.0    &    9.6    & $-$\\
21	&    G304.787$-$00.009    &    4.20 $\pm$ 0.13    &    3.0 $\pm$ 0.10    &    3.0 $\pm$ 0.2    &    0.5$^{\uparrow}$    &    6.5    &    0.5$^{\uparrow}$    & $-$\\
22    &    G304.782$-$00.026    & \multicolumn{2}{c}{unresolved} &    20.5 $\pm$ 0.5    &    21.1 $\pm$ 1.1    &    23.9    &    24.9    &    0.1 \\
23    &    G304.776$-$00.023    & \multicolumn{2}{c}{unresolved} &    2.6 $\pm$ 0.3    &    3.3 $\pm$ 0.2    &    2.9    &    3.5    &    1.0 	\\
24    &    G304.895+00.128    & \multicolumn{2}{c}{unresolved} &    3.2 $\pm$ 0.1    &    2.8 $\pm$ 0.2    &    3.7    &    3.3    &    $-$0.3 	\\
25    &    G304.944+00.411    & \multicolumn{2}{c}{unresolved} &    3.4 $\pm$ 0.3    &    3.2 $\pm$ 0.3    &    3.8    &    3.5    &    $-$0.1 	\\
26$^{\rm \textit{uc}}$    &    G304.930+00.552    &    3.00 $\pm$ 0.53    &    4.2 $\pm$ 0.4   &    2.4 $\pm$ 0.3    &    1.3 $\pm$ 0.1    &    7.3    &    0.8    & $-$\\
27    &    G305.706+00.501    &    \multicolumn{2}{c}{unresolved}  &    3.3 $\pm$ 0.2    &    0.5$^{\uparrow}$     &    4.4    &    0.5$^{\uparrow}$    & $-$3.4  \\
28    &    G305.679+00.019    & \multicolumn{2}{c}{unresolved} &    1.0 $\pm$ 0.1    &    0.8 $\pm$ 0.1    &    1.2    &    0.8    &    $-$0.8 	\\
29    &    G305.714$-$00.102    & \multicolumn{2}{c}{unresolved} &    4.8 $\pm$ 0.1    &    3.4 $\pm$ 0.3    &    6.1    &    4.2    &    $-$0.4 	\\
30    &    G305.698$-$00.135    &    4.20 $\pm$ 4.84    &    6.0 $\pm$ 0.36    &    9.7 $\pm$ 1.4    &    4.8 $\pm$ 0.1    &    30.0    &    14.0    & $-$\\
31    &    G305.663$-$00.150    &    2.17 $\pm$ 0.10    &    1.46 $\pm$ 0.1    &    1.3 $\pm$ 0.1    &    0.8 $\pm$ 0.1    &    2.3    &    0.8    & $-$\\
32    &    G305.729$-$00.176    & \multicolumn{2}{c}{unresolved} &    3.9 $\pm$ 0.2    &    2.7 $\pm$ 0.2    &    4.9    &    3.5    &    $-$0.6 	\\
33    &    G305.672$-$00.191    &    3.00 $\pm$ 0.26    &    4.8 $\pm$ 0.14    &    4.0 $\pm$ 0.3    &    2.0 $\pm$ 0.2    &    6.8    &    3.9    & $-$\\
34    &    G305.672$-$00.223    &  \multicolumn{2}{c}{unresolved}   &    1.9 $\pm$ 0.1    &    0.9 $\pm$ 0.1    &    2.6    &    1.4    & $-$1.8  \\
35    &    G305.477+00.381    & \multicolumn{2}{c}{unresolved} &    0.9 $\pm$ 0.1    &    0.8 $\pm$ 0.1    &    1.0    &    0.8    &    $-$0.1 	\\
36    &    G305.494$-$00.110    & \multicolumn{2}{c}{unresolved}   &    1.1 $\pm$ 0.1    &     0.5$^{\uparrow}$    &    1.9    &     0.5$^{\uparrow}$    & $-$2.4 \\
37$^{\rm \textit{uc}}$    &    G305.200+00.019    &    2.40 $\pm$ 0.16    &    2.4 $\pm$ 0.13    &    7.2 $\pm$ 0.5    &    7.0 $\pm$ 0.8    &    9.2    &    8.6    & $-$\\
38    &    G305.405+00.504    & \multicolumn{2}{c}{unresolved} &    2.2 $\pm$ 0.1    &    1.8 $\pm$ 0.1    &    2.5    &    2.2    &    $-$0.1 \\
39    &    G305.414+00.514    & \multicolumn{2}{c}{unresolved} &    1.3 $\pm$ 0.1    &    0.8 $\pm$ 0.1    &    1.4    &    0.8    &    $-$1.2 	\\
40    &    G305.752+00.410    & \multicolumn{2}{c}{unresolved} &    1.3 $\pm$ 0.1    &    0.8 $\pm$ 0.1    &    1.7    &    0.9    &    $-$1.3 	\\
41    &    G305.132+00.502    & \multicolumn{2}{c}{unresolved} &    1.6 $\pm$ 0.1    &    1.7 $\pm$ 0.1    &    1.8    &    2.0    &    0.2 \\
42    &    G304.843+00.085    & \multicolumn{2}{c}{unresolved} &    1.9 $\pm$ 0.1    &    0.5$^{\uparrow}$    &    2.1    &    0.5$^{\uparrow}$    &    $-$0.2 \\
43    &    G304.861+00.051    & \multicolumn{2}{c}{unresolved} &    2.4 $\pm$ 0.1    &    1.2 $\pm$ 0.1    &    3.1    &    1.8    &    $-$1.0 	\\
44    &    G305.389+00.235    & \multicolumn{2}{c}{unresolved} &    1.3 $\pm$ 0.1    &    0.9 $\pm$ 0.1    &    1.5    &    1.4    &    $-$0.1 	\\
45    &    G305.572+00.264    & \multicolumn{2}{c}{unresolved} &    0.8 $\pm$ 0.1    &    0.5 $\pm$ 0.2    &    0.8    &    0.7    &    $-$0.1 	\\
46    &    G305.694$-$00.322    & \multicolumn{2}{c}{unresolved} &    3.0 $\pm$ 0.1    &    1.5 $\pm$ 0.1    &    3.9    &    1.8    &    $-$1.0 	\\
47    &    G305.410$-$00.287    & \multicolumn{2}{c}{unresolved} &    0.8 $\pm$ 0.1    &    0.8 $\pm$ 0.1    &    0.7    &    0.8    &    $-$0.3 	\\
       \hline
       \end{tabular}
\end{table*}

\begin{table*}
\centering     
  \begin{tabular}{cccccccccccc}
   \hline 
   Source  & Source	 & \multicolumn{2}{c}{Dimensions}  & \multicolumn{4}{c}{Observed flux density} &Spectral \\ 
	       Index   & Name  & \multicolumn{2}{c}{Observed} & \multicolumn{2}{c}{Peak}  &\multicolumn{2}{c}{Integrated}& Index \\ \cline{3-4} \cline{5-8}
	       	       &          & maj & min &   \multicolumn{2}{c}{(mJy beam$^{-1}$)}       & \multicolumn {2}{c}{(mJy)} &($\alpha$) \\
	        &           & ('')   & ('')    & \textit{f}$_{5.5}$ & \textit{f}$_{8.8}$ & \textit{f}$_{5.5}$ & \textit{f}$_{8.8}$ &    \\
	        \hline

48    &    G305.198$-$00.271    & \multicolumn{2}{c}{unresolved} &    1.1 $\pm$ 0.1    &    0.9 $\pm$ 0.1    &    1.4    &    1.2    &    $-$0.1 	\\
49    &    G304.845$-$00.216    & \multicolumn{2}{c}{unresolved} &    2.1 $\pm$ 0.3    &    1.2 $\pm$ 0.3    &    2.6    &    1.4    &    $-$1.0 	\\
50    &    G304.819$-$00.152    & \multicolumn{2}{c}{unresolved} &    1.1 $\pm$ 0.2    &    1.3 $\pm$ 0.2    &    1.3    &    1.6    &    0.4 	\\
51    &    G304.803+00.222    & \multicolumn{2}{c}{unresolved} &    1.7 $\pm$ 0.1    &    2.0 $\pm$ 0.2    &    2.1    &    2.5    &    0.6 	\\
52    &    G305.669+00.428    &   \multicolumn{2}{c}{unresolved}     &    1.0 $\pm$ 0.3    &   0.5$^{\uparrow}$    &    1.5    &    0.5$^{\uparrow}$    & $-$1.2 \\
53    &    G305.714+00.353    & \multicolumn{2}{c}{unresolved} &    0.8 $\pm$ 0.1    &    0.5$^{\uparrow}$    &    0.9    &    0.5$^{\uparrow}$     &    $-$1.3 \\
54    &    G305.200+00.448    & \multicolumn{2}{c}{unresolved} &    0.8 $\pm$ 0.3    &    0.5$^{\uparrow}$    &    1.1    &    0.5$^{\uparrow}$    &    $-$1.4\\
55    &    G304.851+00.502    & \multicolumn{2}{c}{unresolved} &    0.9 $\pm$ 0.2    &   0.5$^{\uparrow}$   &    0.7    &    0.5$^{\uparrow}$    &    $-$0.4 \\
56$^{\rm \textit{st}}$    &    G305.302+00.052    & \multicolumn{2}{c}{unresolved} &    1.1 $\pm$ 0.2    &    1.2 $\pm$ 0.2    &    1.3    &    1.5    &    0.1 \\
57    &    G304.818+00.235    &     \multicolumn{2}{c}{unresolved}   &    0.8 $\pm$ 0.1    &    0.9 $\pm$ 0.1    &    0.8    &    0.9    & 0.2 \\
58    &    G305.291$-$00.094    &     \multicolumn{2}{c}{unresolved}  &    0.5 $\pm$ 0.1    &    0.5$^{\uparrow}$     &    0.6    &    0.5$^{\uparrow}$    & $-$0.8 \\
59$^{\rm \textit{st}}$    &    G305.335+00.074    & \multicolumn{2}{c}{unresolved} &    1.5 $\pm$ 0.1    &    1.5 $\pm$ 0.1    &    1.7    &    2.2    &    0.5 \\
60$^{\rm \textit{st}}$    &    G305.522+00.063    & \multicolumn{2}{c}{unresolved} &    0.8 $\pm$ 0.1    &    0.5$^{\uparrow}$     &    1.0    &    0.5$^{\uparrow}$     &    $-$0.6\\
61    &    G305.499+00.050    & \multicolumn{2}{c}{unresolved} &    0.5 $\pm$ 0.1    &    0.5$^{\uparrow}$     &    0.6    &     0.5$^{\uparrow}$    &    $-$0.8 \\
62    &    G305.385$-$00.147    &    \multicolumn{2}{c}{unresolved}     &    0.5 $\pm$ 0.2    &    0.5$^{\uparrow}$     &    0.7    &    0.5$^{\uparrow}$   & $-$0.3  \\
63$^{\rm \textit{\textit{st}}}$   &    G305.091$-$00.095    & \multicolumn{2}{c}{unresolved} &    0.5 $\pm$ 0.1    &    0.5$^{\uparrow}$     &    0.5    &    0.5$^{\uparrow}$     &    0.1 \\
64    &    G305.047$-$00.120    & \multicolumn{2}{c}{unresolved} &    0.5 $\pm$ 0.1    &    0.5 $\pm$ 0.1    &    0.6    &    0.7    &    $-$0.1 	\\
65   &    G305.558+00.459    &    \multicolumn{2}{c}{unresolved} &    0.6 $\pm$ 0.1    &    0.5$^{\uparrow}$    &    0.7    &    0.5$^{\uparrow}$    & $-$0.8\\
66    &    G305.076+00.396    & \multicolumn{2}{c}{unresolved} &    0.7 $\pm$ 0.1    &    0.6 $\pm$ 0.1    &    0.8    &    0.7   &    $-$0.8 	\\
67$^{\rm \textit{st}}$    &    G305.631+00.000    & \multicolumn{2}{c}{unresolved} &    0.9 $\pm$ 0.1    &    1.2 $\pm$ 0.2    &    1.1    &    1.5    &    0.7 \\
68    &    G305.466+00.374    & \multicolumn{2}{c}{unresolved}     &    0.6 $\pm$ 0.3    &    0.6 $\pm$ 0.1    &    1.1    &    1.1    & $-$0.5 \\
69    &    G305.284+00.365    &   \multicolumn{2}{c}{unresolved}     &    0.5 $\pm$ 0.1    &    0.5$^{\uparrow}$     &    0.6    &    0.5$^{\uparrow}$    & $-$0.2 \\
70    &    G305.596$-$00.211    & \multicolumn{2}{c}{unresolved} &    0.6 $\pm$ 0.2    &   0.5$^{\uparrow}$     &    0.8    &    0.5$^{\uparrow}$     &    $-$0.2 \\
       \hline
       \end{tabular}
        \caption{Identifiers and observed properties of the 71 detected compact radio sources. For sources that fall below the detection limit at 8.8\,GHz we present the upper limit $3 \times \sigma$ value where $\sigma$ is the rms noise. Superscript above the source index indicates the nature of the source as derived in Section\,3, \textit{st} corresponds to stellar, \textit{uc} to \mbox{UC HII} candidate and \textit{brc} to bright$-$rimmed cloud sources with no superscript are background.The spectral index is derived from the 1\,GHz split dataset and presented for unresolved sources only.}
\end{table*}


To explore the nature of the detected radio emission we extract images from the Spitzer telescope Galactic Legacy Infrared Mid-Plane Survey Extraordinaire (GLIMPSE; \citealt{Benjamin2003, Churchwell2009}). The GLIMPSE survey utilises the InfraRed Array Camera (IRAC; \citealt{Fazio2004}) on Spitzer to map emission at 3.6, 4.5, 5.8 and 8.0\,$\umu$m with a spatial resolution of $\sim 2\arcsec$ FWHM well matched to the observations presented here. The four IRAC bands trace different physical properties and so analysis of these provides a means to infer the origin of the radio emission. The 3.6\,$\umu$m band is dominated by field stars, while an excess in the 4.5\,$\umu$m band is thought to arise from shocked molecular hydrogen and CO bandheads, possibly indicating the presence of an outflow. The 5.8 and 8.0\,$\umu$m IRAC bands are dominated by strong polycyclic aromatic hydrocarbon (PAH) features that are excited in the photo-dominated regions (PDRs) that lie in a thin shell of neutral gas located between an ionisation front and cold molecular gas.

\begin{figure*}
\includegraphics[width=0.498\textwidth, trim=20 20 60 40]{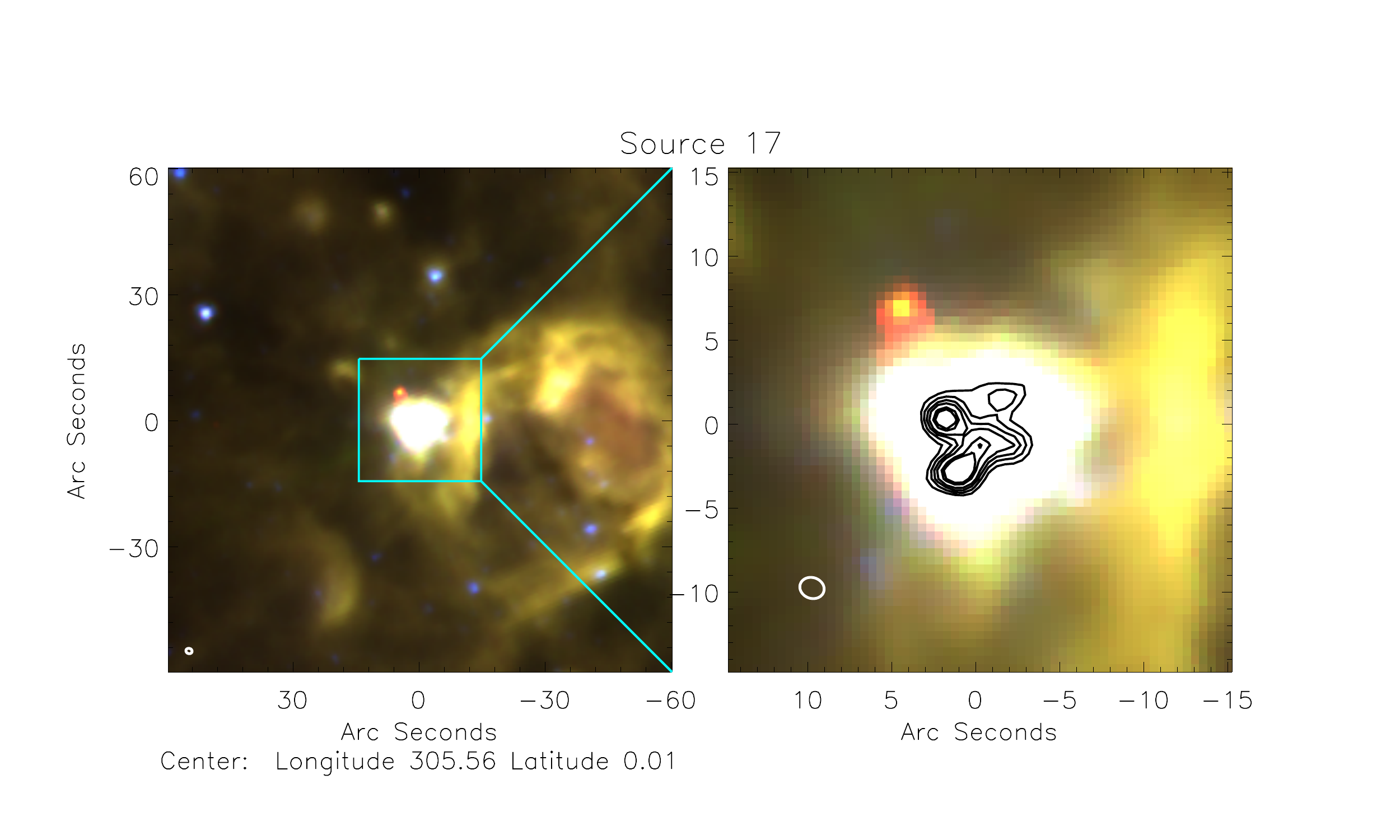} 
\includegraphics[width=0.498\textwidth, trim=20 20 60 40]{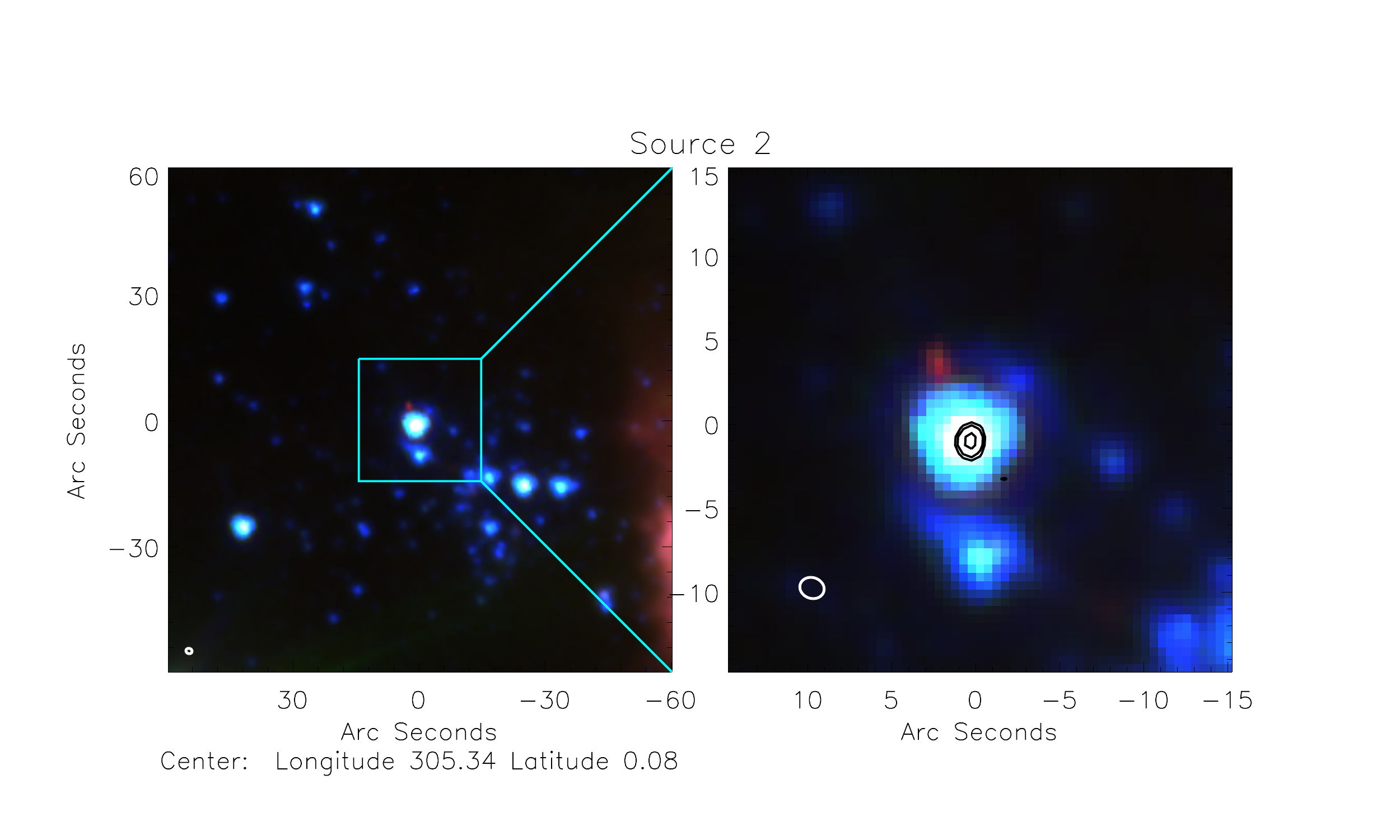} 
\includegraphics[width=0.498\textwidth, trim=20 20 60 40]{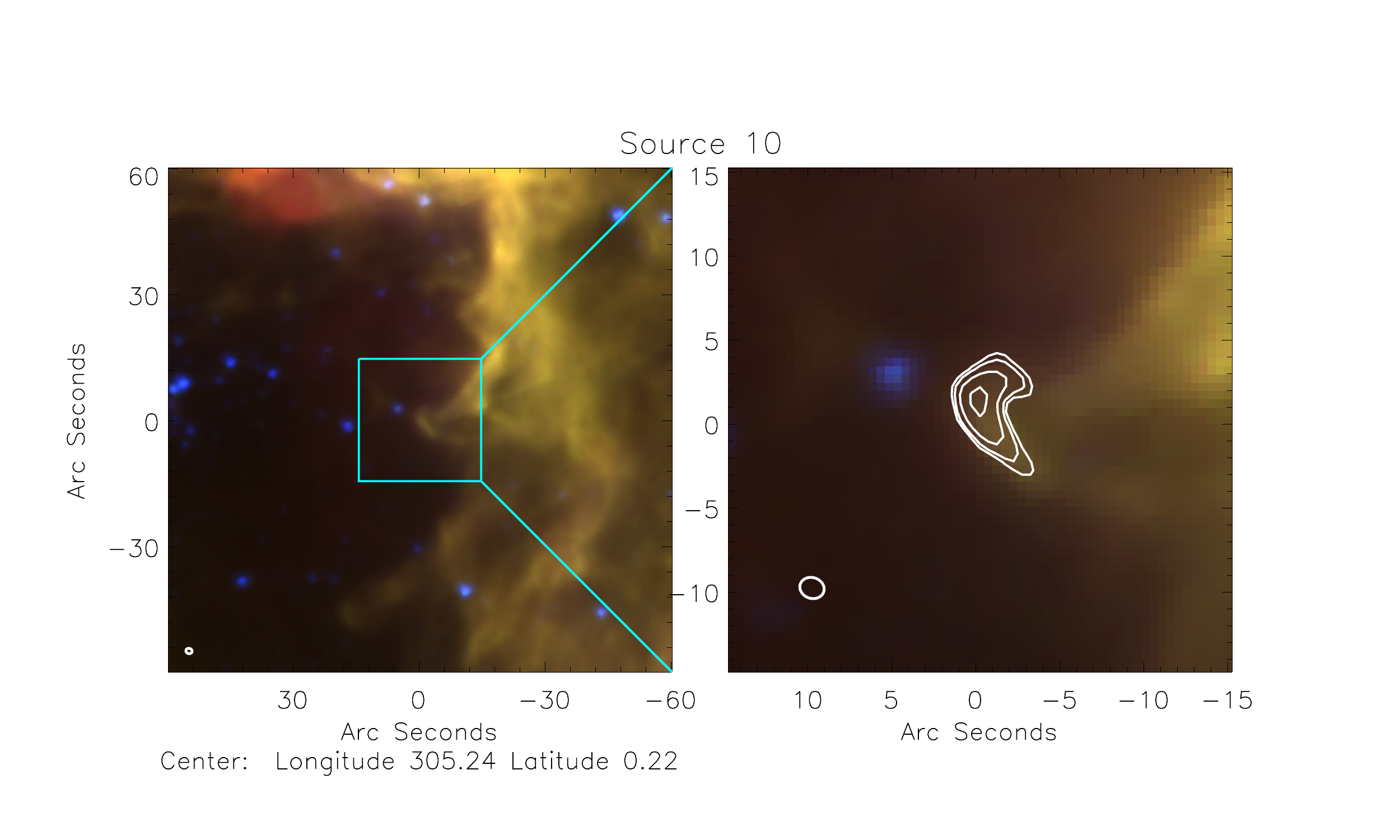} 
\includegraphics[width=0.498\textwidth, trim=20 20 60 40]{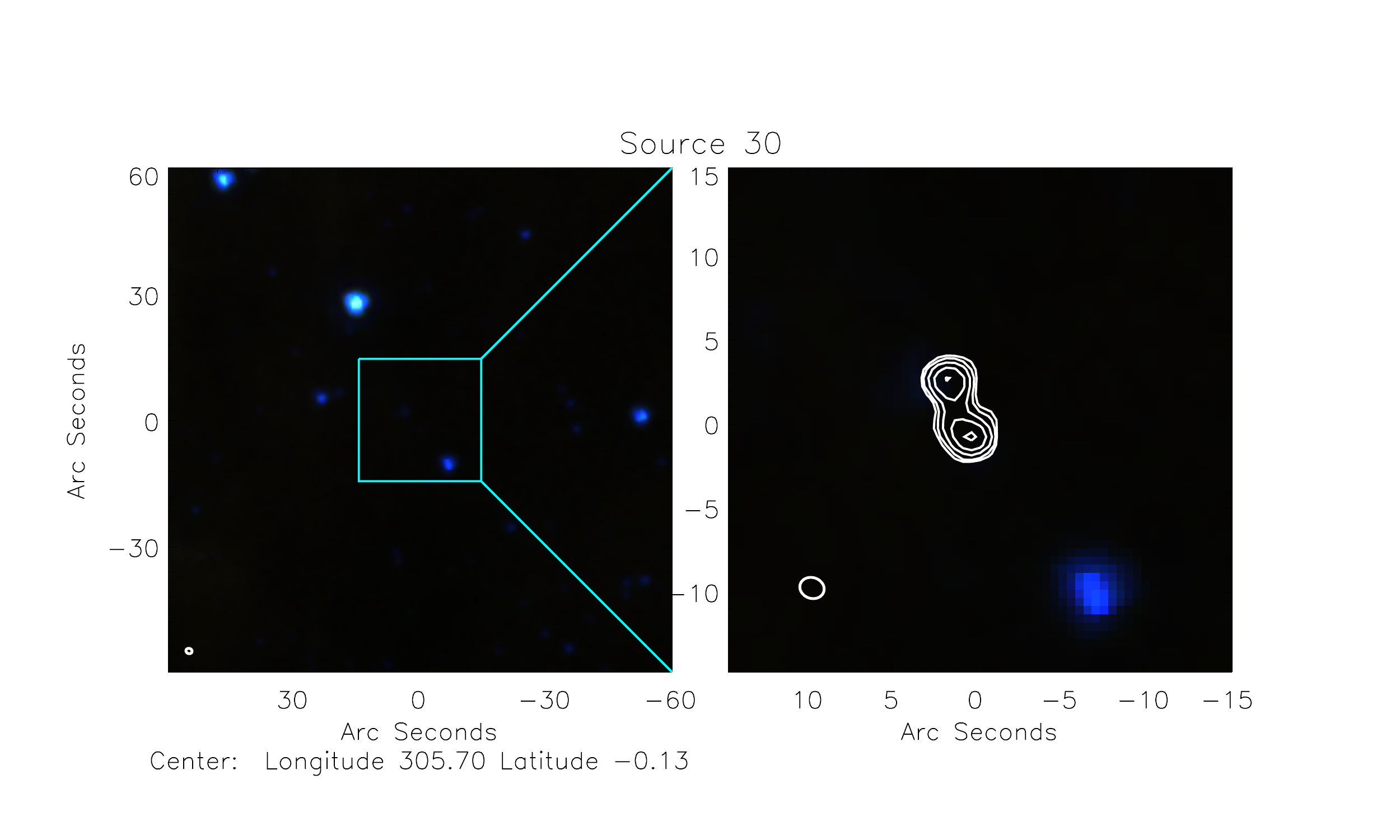} 
\caption{Three colour 4.5, 5.8 and 8.0\,$\umu$m composite GLIMPSE (b, g, r respectively) cut-out images centred on the peak 5.5\,GHz radio emission with contours beginning and incrementing by $3\sigma$. An example of a \mbox{UC\,HII} region (top left), stellar (top right), BRC (bottom left) and background galaxy (bottom right) are shown. These are classified following the criteria outlined in Section\,3. The beam is presented in the lower left corner of each image.}
\label{IRUContaminate_fig}
\end{figure*}

To take advantage of the sensitivity to different physical phenomenon we have created false three-colour images, centred on the 5.5\,GHz peak, using the 4.5, 5.8 and 8.0\,$\umu$m IRAC bands in blue, green and red, respectively (cf. \citealt{Cohen2007,Urquhart2009}). From a visual examination of these images, we find that the radio sources can be separated into four distinct categories. In Fig.\,2 we present an example of each of these and describe the main features of each category below:

\begin{itemize}
\item Radio emission is coincident with strong compact mid-infrared emission, which is itself associated with extended PAH emission and/or extinction, which are commonly associated with star formation regions. These are considered \mbox{UC\,HII} candidates (Fig.\,2 top-left).
\item Radio and mid-infrared emission are both unresolved and correlated and the mid-infrared emission is dominated by the shorter (blue) wavelengths. We consider that the radio emission seen towards these sources arises from a stellar source. Additionally, these radio and mid-infrared sources are often isolated which suggests they are not associated with star formation. (Fig.\,\ref{IRUContaminate_fig} top-right).
\item We classify a source as a bright-rimmed cloud (BRC) if the radio and mid-infrared emission have a similar cometary morphology with no obvious core and the radio emission is offset from the mid-infrared emission (Fig.\,2 bottom-left).
\item Radio sources that are devoid of any mid-infrared emission are assumed extragalactic radio sources. A large proportion of the unresolved radio sources fall into this category (Fig.\,2 bottom-right).
\end{itemize}

We derive the spectral index ($\alpha$) of sources by taking the logarithm and fitting a straight line to the relation $S_{\rm int}\propto \nu^{\alpha}$ where $S_{\rm int}$ is the integrated flux and $\nu$ is the frequency. To search for non-linear variation of flux with frequency we split the full 2\,GHz bandwidth into 1\,GHz wide bands centered at 5.0, 6.0, 8.3 and 9.3\,GHz and generate images following the steps outlined in Section\,2.2. Given the errors we find no convincing evidence of non-linear SEDs. As mentioned in Section\,2.2 the integrated flux for resolved sources is a lower limit and cannot be reliably estimated in addition the $uv$ plane has been sampled differently at the two frequencies, as the ATCA is not a scaled array, and so the flux cannot be reliably compared. For these reasons the spectral index cannot be determined reliably for resolved sources and is presented for only the unresolved and isolated sources. 

We will further consider the \mbox{UC\,HII} region candidates in the next section. For the remainder of this section we briefly discuss the other classes of object discovered.

We find eight of the 71 radio sources (Fig.\,\ref{IRUContaminate_fig}) are associated with more evolved stars (isolated, unresolved and possessing blue colours). Approximately, half of these stellar sources are found towards the centre of G305 and are associated with well known evolved massive stars. Two sources, 2 and 59, are associated with the most massive members of Danks\.1 (D1-1, D1-2), whilst Source\,3 is associated with the Wolf-Rayet star WR\,48A and Source\,56 with the massive star MDM3 (Paper\,2). Examination of the spectral index reveals the radio emission associated with six stellar sources is positive (2, 3, 56, 59, 63 67) and only two are negative (11, 60). Sources with a positive spectral index are most likely generating thermal emission associated with an ionised stellar wind whilst negative spectral index sources are most likely generating non-thermal emission via wind shocks or colliding winds from binary stars \citep{Dough2003}. Further analysis of these sources is beyond the scope of this study.

Bright-rimmed clouds are found on the edge of evolved HII regions where the ionising radiation emitted from an OB star(s) ionises the surface of a molecular cloud \citep{Sugitani1991,Thompson2004}. This molecular material is often swept into a cometary morphology by the radiation pressure and results in a dense core at the head that shields the remaining cloud material resulting in a finger or column. Comparison with mid-infrared data reveals Source\,10 has this characteristic morphology and is found on the inner borders of a shell of diffuse 5.8\,$\umu$m emission that has been identified as an evolved HII region \citep{Caswell1987, Clark2004}. The cometary morphology of the radio and mid-infrared emission points to an ionising source to the east of the BRC towards the center of the evolved HII region. Compelling candidates for the source of ionisation are found within the HII region in the massive star L05-A2 (Paper\,2, Table\,2) and a deeply embedded IR excess cluster \citep{Longmore2007}. 

As previously mentioned we failed to associate the majority of radio sources with any mid-infrared emission and these are assumed extragalactic background sources. To test this assumption it is useful to compare the number in this category with the estimated number of background radio sources $(\left \langle N \right \rangle)$ one would expect given the size of the observed region and the frequency at which the observations were performed. We can estimate this number empirically from extragalactic source counts using the following equation from \citet{Anglada1998}:

\begin{equation}
\left \langle N \right \rangle\simeq\left (\frac{\theta_{\rm f}}{\theta_{\rm A}}\right )1.1 S_{\rm {0}}^{-0.75}
\label{eqn:anglada}
\end{equation}

\noindent where $\theta_{\rm f}$ is the diameter of the observed field, $\theta_{\rm A}$ is the FWHM of the primary beam in arc-minutes and $S_{\rm{0}}$ is the sensitivity at 5.5\,GHz. Using this equation we estimate that $\sim\,60\pm 8$ background sources should be detected in the $\sim\, 1.0 \times 1.0 \degr$ field. From our comparison with the mid-infrared we find 56 compact radio sources that fit the criteria of extragalactic sources, which is in good agreement with the result obtained above from Equation \ref{eqn:anglada}. These background sources are mainly unresolved point sources but there are a number of resolved sources that exhibit lobes of emission similar to the radio lobes seen in active radio galaxies (Fig.\,2 bottom-right). We expect to find a negative or flat spectral index for background radio galaxies associated with non-thermal radiation and we find this is the case. Only 9 of these background sources have $\alpha > -0.1$ consistent with optically thick thermal free-free emission while 42 have $\alpha < -0.1$ indicating a non-thermal emission mechanism.

\begin{figure*}
\includegraphics[width=0.498\textwidth, trim=20 20 60 40]{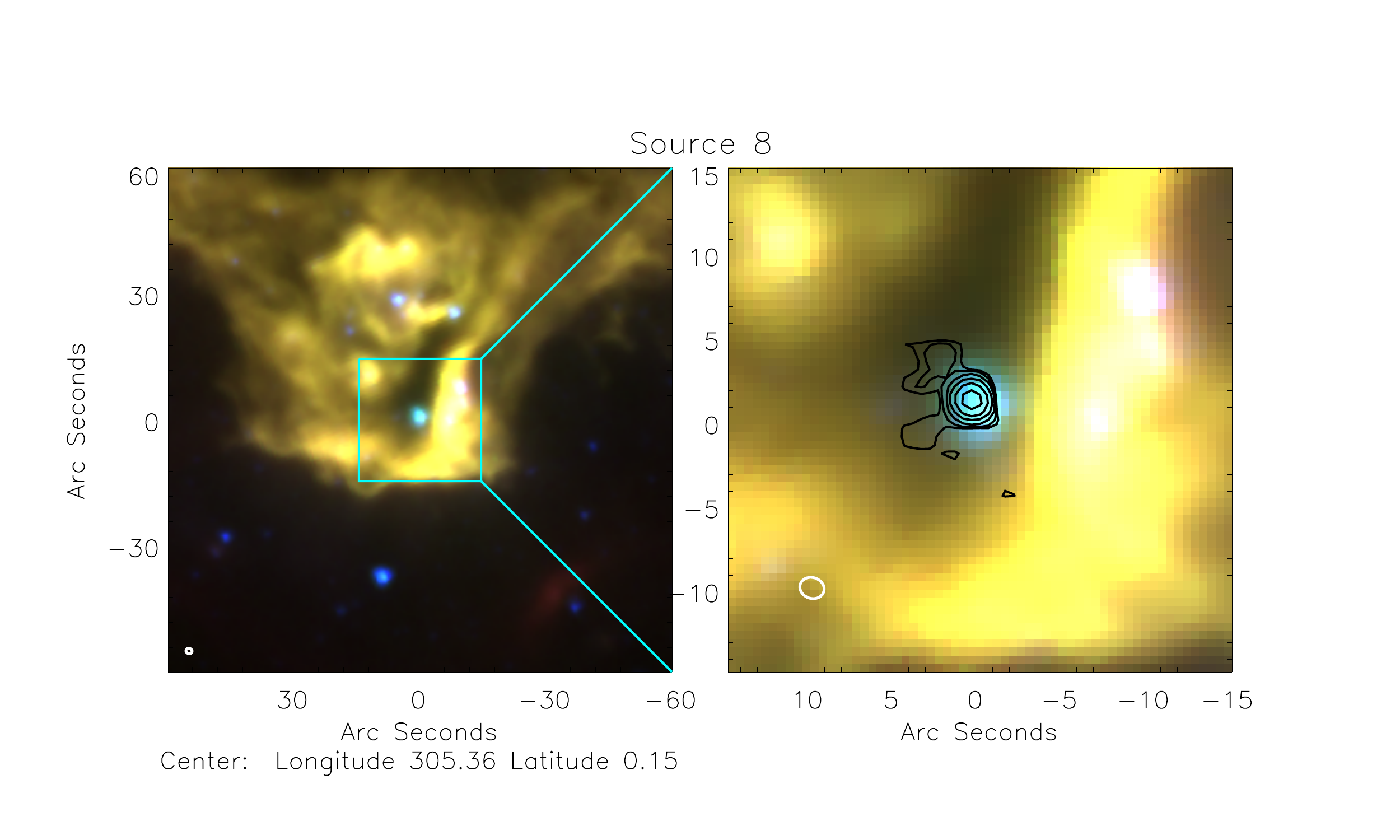} 
\includegraphics[width=0.498\textwidth, trim=20 20 60 40]{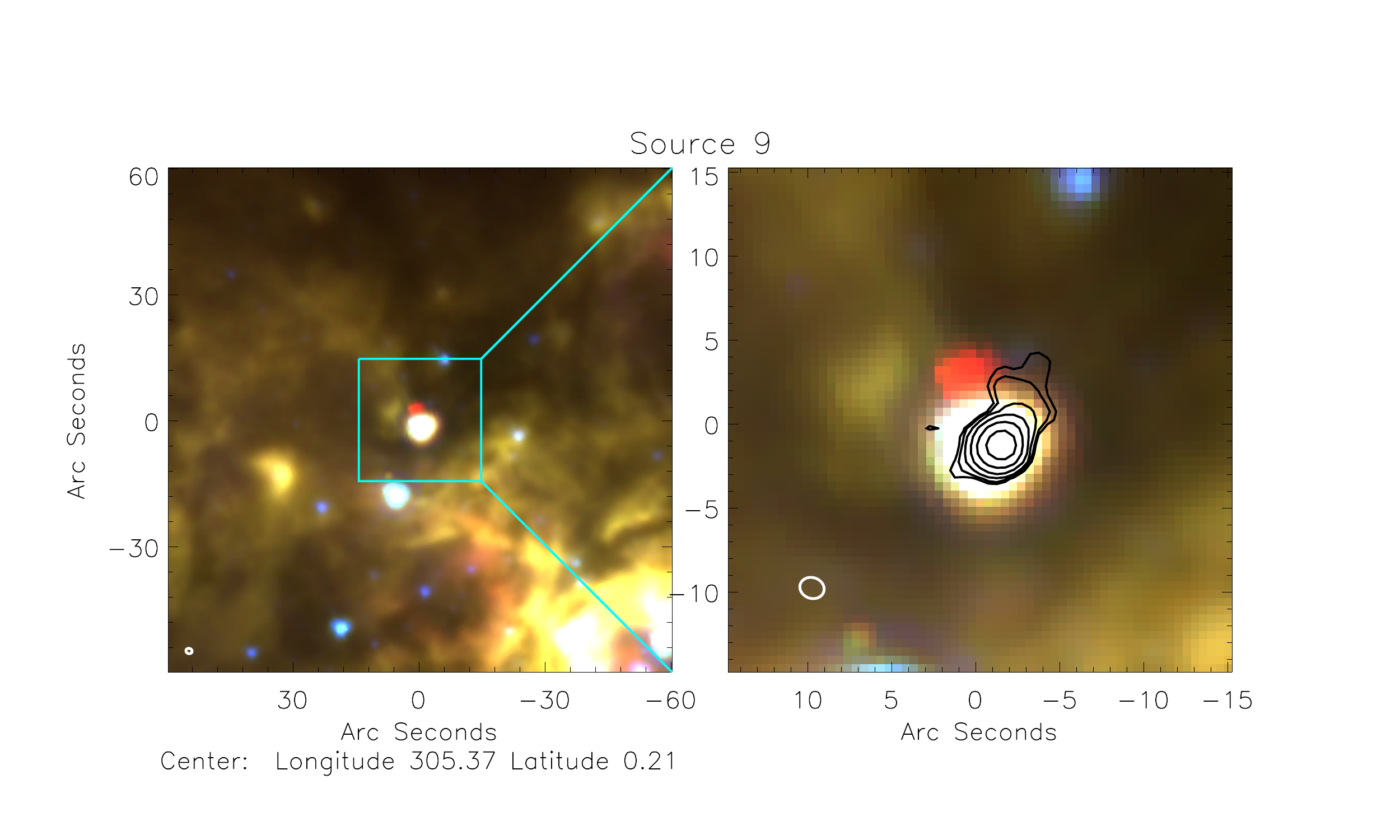} 
\includegraphics[width=0.498\textwidth, trim=20 20 60 40]{G305-CompactSource-GLIMPSE-Comb-5500-17} 
\includegraphics[width=0.498\textwidth, trim=20 20 60 40]{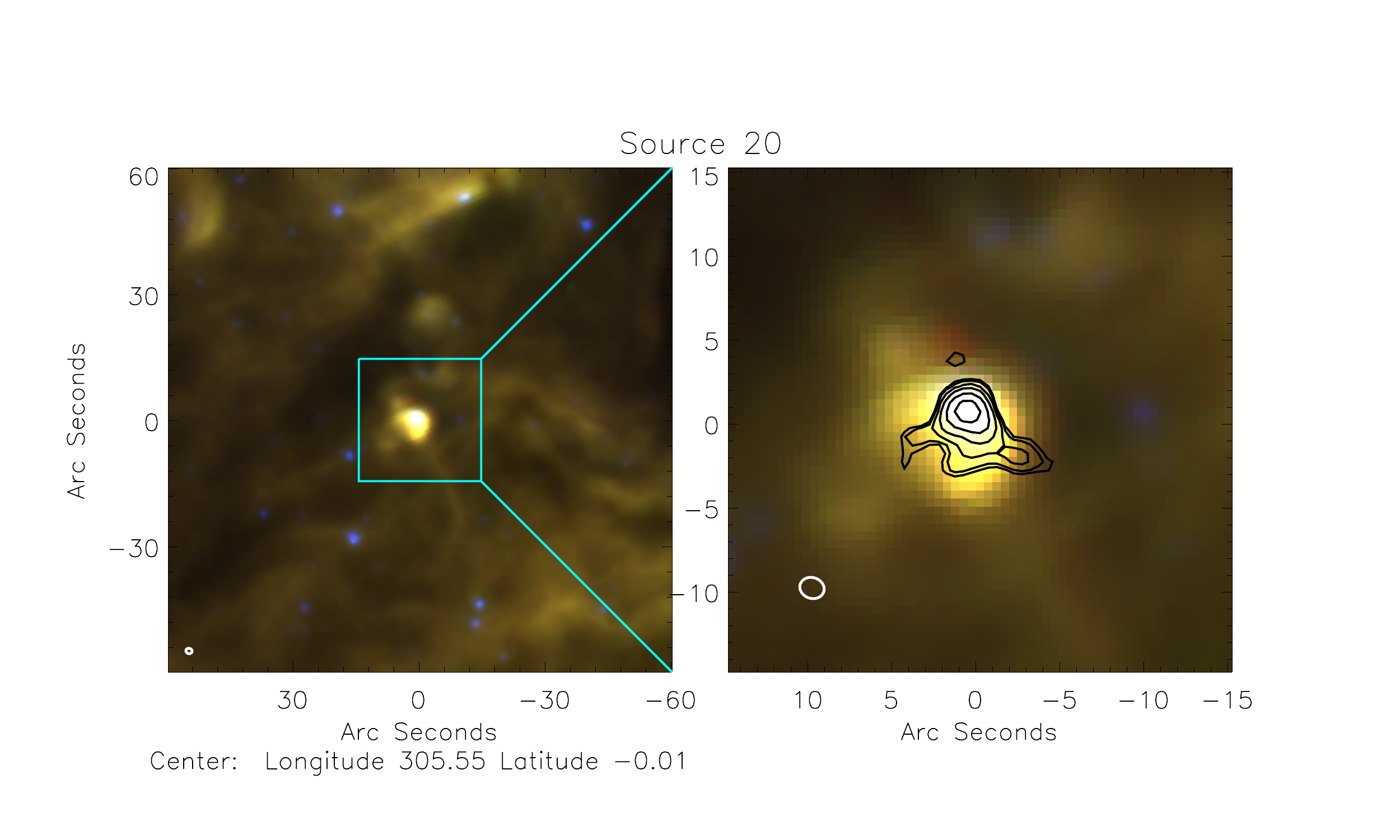} 
\includegraphics[width=0.498\textwidth, trim=20 20 60 40]{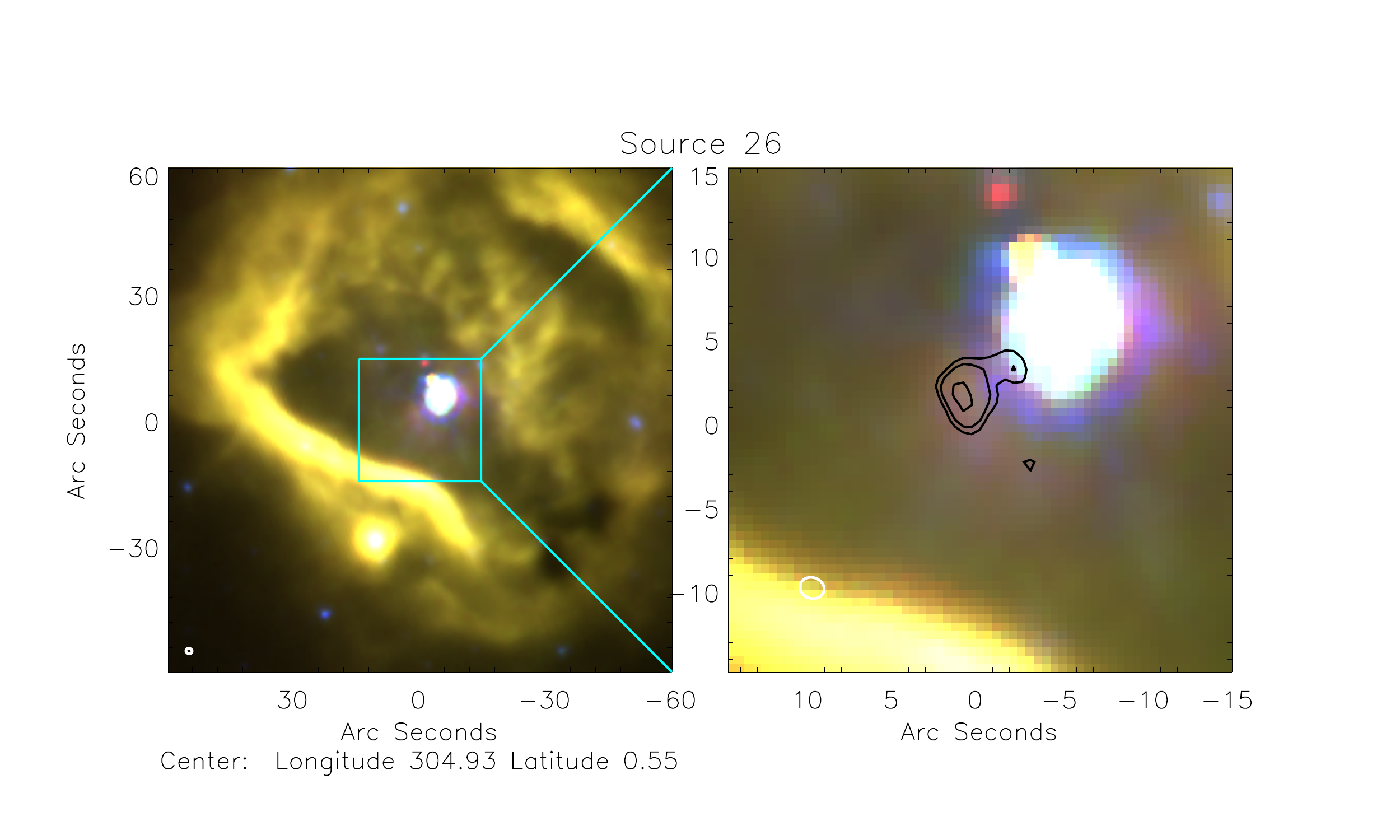} 
\includegraphics[width=0.498\textwidth, trim=20 20 60 40]{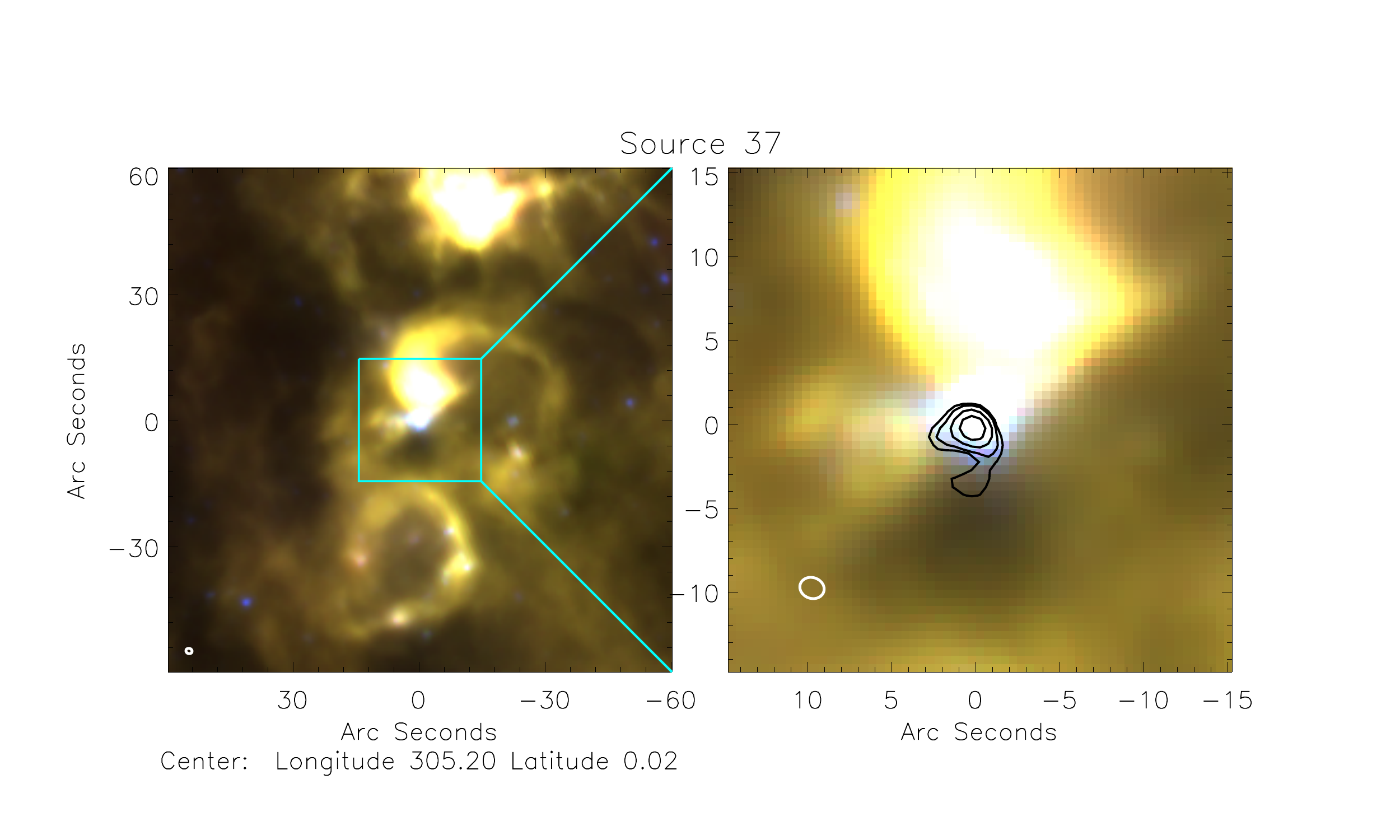} 
\caption{Three colour 4.5, 5.8 and 8.0\,$\umu$m composite GLIMPSE (b, g, r respectively) cut-out images centred on the peak 5.5\,GHz radio emission of the six candidate \mbox{UC\,HII} regions. First panel shows a wide field of view image with a cyan box centered on the peak of the radio emission. Second panel shows the outlined field of view the cyan box. Black contours are of the 5.5\,GHz radio emission with contours beginning and incrementing by $3\sigma$. Sources are coincident with mid-infrared emission indicative of an embedded stellar source (see criteria in Section\,3). The beam is presented in the lower left corner of each image.}
\label{IRUCompact_fig}
\end{figure*}

\section{Physical Parameters of \mbox{UC\,HII} regions}

In the previous section, we described how mid-infrared data was used to separate the compact radio sources into four distinct categories. We identified six sources that appear to be good candidates for classification as \mbox{UC\,HII} regions (see Fig.\,3 for mid-infrared three-colour images of these). In this section, we will determine the physical properties of these candidate \mbox{UC\,HII} for comparison with bona fide \mbox{UC\,HII} regions reported in the literature.

In the following analysis, we assume the origin of the radio emission is free-free (bremsstrahlung), optically thin and thermal arising in an ionised circumstellar environment. This assumption must be treated with caution as the opacity of a typical \mbox{UC\,HII} region with emission measure $> 10^7\rm\, pc\, cm^{-6}$, diameter of $\sim 0.1$\,pc and electron density $> 10^{4} \rm\, cm^{-3} $ is found to turn over from optically thin to optically thick below 5\,GHz \citep{Kurtz2005A}. To test the validity of this assumption we derive the physical properties using both the 5.5 and 8.8\,GHz data and find they agree to within a factor of two, which, would suggest that the assumption of optically thin emission at 5.5\,GHz is valid. We therefore only report the properties derived from the higher signal to noise 5.5\,GHz data. 

As mentioned in Section\,2.2 the integrated flux of resolved compact radio sources is underestimated due to the removal of short baselines used to emphasise small-scale structure. The peak emission of sources is unaffected by the removal of short baselines and so we derive the peak (mJy beam$^{-1}$) properties, averaged over the area of the synthesised beam, following the method of WC89A. The beam brightness temperature is estimated from:

\begin{equation}
T_{\rm{b}}=\frac{S_{\nu}10^{-29}c^2}{2\nu^2k\Omega_{\rm{b}} } \; [\rm K]
\end{equation}

\noindent where $S_{\nu}$ is the peak flux density (mJy beam$^{-1}$), $\nu$ is the frequency (Hz), $c$ is the speed of light (m\,s$^{-1}$) and $\Omega_{\rm{b}}$ is the beam solid angle ($3.36\times 10^{-11}$ sr at 5.5\,GHz). The peak optical depth $\tau$ is estimated using $T_{\rm{b}}=T_{\rm{e}}(1-e^{-\tau})$ assuming the beam is uniformly filled with ionised gas. The electron temperature of the ionised gas within HII regions has been shown to vary between 6800-13000\,K \citep{Spitzer1950,Caswell1987}. The electron temperature towards regions in G305 is unknown and so we assume a value of $T_{\rm{e}}=10^{4}$\,K throughout the rest of this analysis. This assumption results in an uncertainty of $\sim 20\%$ in the derived optical depth and $<10\%$ in the physical properties derived below. The optical depth $\tau$ may be written:

\begin{equation}
\tau =-ln\left ( 1-\frac{T_{\rm{b}}}{10^{4}} \right )
\end{equation}

\noindent We calculate the peak emission measure $EM$ from the expression of the optical depth $\tau$ for free-free radiation:

\begin{equation}
EM=\frac{\tau}{8.235\times10^{-2} T_{\rm{e}}^{-1.35}\nu^{-2.1} }\; [\rm{pc\,cm^{-6}]}
\end{equation}

\noindent where the frequency is expressed in GHz. The electron density ($n_{\rm{e}}$) is given by:

\begin{equation}
n_{\rm{e}}=\sqrt{\frac{EM}{\Delta s}} \; [\rm cm^{-3}]
\end{equation}

\noindent where $\Delta s$ is the optical path length through the peak and is defined as the geometrical average of the two axes of the source after applying the correction of \citet{Panagia1978}. We estimate the peak ionised gas mass ($M_{\rm{HII}}$) by multiplying the peak electron density by the proton mass and volume of the emitting region assuming spherical morphology. 

The remaining physical properties ($N_{\rm Ly}$, M$_*$ and spectral type) have been derived using the integrated flux and it is important to reiterate that the since the integrated flux is underestimated these properties should be considered lower limits. The total ionising photon flux of the Lyman continuum ($N_{\rm{Ly}}$) is independent of source geometry and is determined using the modified equation (7) presented in \citet{Carpenter1990}:

\begin{equation}
N_{\rm{Ly}} =7.7\times10^{43}S_{\rm\,int}D^2\nu^{0.1} \; [\rm s^{-1}]
\end{equation}

\noindent where $N_{\rm{Ly}}$ is the total number of Lyman photons emitted per second, $S_{\rm int}$ is the integrated radio flux (mJy), $D$ is the distance to the source ($3.8\pm0.6$\,kpc) and $\nu$ is the frequency of the observation (GHz). This value of the ionising flux should be considered a lower limit due to dust absorption of the UV flux and the under estimated integrated flux. 

We are able to estimate the mass of the ionising source (M$_*$) responsible for the observed Lyman flux by interpolating between the calculated values of mass and Lyman flux presented in Table\,1 of \citet{Davies2011}. The values in this table were computed by combining the results from stellar atmosphere models by various authors, see \citet{Davies2011} for more details. The resultant masses are presented in Col.\,10 of Table\,3 and should be treated as a lower limit with errors of $\sim30-50\%$ caused by the calibration of Lyman flux to ionising source mass \citep{Davies2011}.

We estimate the spectral type of the ionising star by comparing the estimated Lyman continuum flux to the derived value of the total number of ionising photons generated by massive stars tabulated by \citet{Panagia1973}. We assume the radio emission observed is caused by a single zero age main sequence (ZAMS) star, assuming no UV flux is absorbed by dust and that the HII region is ionisation bounded. Whilst the most massive star will dominate, the Lyman flux if there are multiple stars that make a significant contribution to the ionising flux the spectral type will be later than we have estimated. Conversely, if there is significant absorption by dust and/or the nebular is not ionisation bounded the spectral type may be earlier than our estimate.

\begin{table*}
\begin{tabular}{cccccccc|ccccccc}
 \hline Source& Source	&Radius & T$_{\rm b}$ & $\tau$ & n$_{\rm e}$ & EM & M$\rm _{\rm HII}$ & Log N$_{\rm Ly}$ & M$_*$ & Spectral\\
 	Index& Name & (pc) & (K) &(10$^{-3}$) & (10$^4$\,cm$^{-3}$) & (10$^6$\,pc cm$^{-6}$)& (10$^{-3}$\msun)& (s$^{-1} $)& (\msun) &Type\\
	 \hline
8&G305.362+00.150&0.04&770.0&80.03&1.03&8.76&74.0&46.62&13.8&B0.5	\\
9&G305.368+00.213&0.08&900.0&94.02&0.79&10.3&440.0&46.95&14.8&B0.5	\\
17&G305.562+00.013&0.11&240.0&23.80&0.34&2.60&480.0&46.65&13.9&B0.5	\\
20 &G305.553$-$00.012&0.08&270.0&27.29&0.44&2.99&200.0&46.36&13.0&B0.5	\\
26&G304.930+00.552&0.07&77.0&7.72&0.25&0.84&84.0&46.02&12.0&B1	\\
37&G305.200+00.019&0.04&230.0&23.34&0.58&2.56&32.0&46.12&12.3&B1	\\
\hline
 \end{tabular}
 \protect\footnotetext[1]{table footnote 1}
 \caption{Derived source properties for candidate \mbox{UC\,HII} regions.}
 \label{UCProperties_table}
 \end{table*}	

The physical properties determined for all but one of the candidate \mbox{UC\,HII} regions in Table\,\ref{UCProperties_table} are typical of \mbox{UC\,HII} regions described in the literature e.g.\ small diameter $(< 0.1$\,pc), high electron density ($> 10^{4} \rm\, cm^{-3} $) and high emission measure ($> 10^7\rm\, pc\, cm^{-6}$). We find that the derived physical properties for Source\,26 are significantly lower than the other candidate \mbox{UC\,HII} regions. Moreover, comparison with the \NH\ emission maps (Paper\,1) reveals a lack of dense gas towards Source\,26 that would indicate that the region is either a more evolved HII region having already dissipated its natal cloud or has another origin. Visual examination of the mid-infrared image of this source, presented in Fig.\,3, reveal that the radio emission is actually offset from the bright mid-infrared source that dominates the image. It is possible that this radio emission is in fact another extragalactic background source and we have therefore removed Source\,26 from the list of \mbox{UC\,HII} regions.
 
For the five remaining sources we find radii ranging from 0.04-0.11\,pc, electron densities in the range of 0.34-$1.03\times10^4$\,cm$^{-3}$ and emission measures between 2.56-$10.3\times10^6$\,pc\,cm$^{-6}$. These values correspond to \mbox{UC\,HII} regions in the classification scheme of \citet{Kurtz2002} and WC89A, B. Based on the correlation with mid-infrared emission and derived physical properties we classify the compact radio sources 8, 9, 17, 20 and 37 as \mbox{UC\,HII} regions.

\section{Discussion}

In this section we will look at the distribution of \mbox{UC\,HII} regions found towards G305 with respect to the overall structure of the complex and the other star formation tracers that have been reported in the literature. To help facilitate the following discussion we present a GLIMPSE 5.8\,$\umu$m image of the G305 complex in Fig.\,\ref{Multi}. On this image we show the distribution of \mbox{UC HII} regions, dense gas as traced by the NH$_3$ (1,1) and (3,3) inversion transition and water masers presented in Paper\,1, methanol masers identified by the Methanol Multibeam (MMB) survey \citep{Green2009} and massive young stars identified by the Red MSX Source (RMS; \citealt{Urquhart2008}) survey.

\subsection{Distribution of \mbox{UC\,HII} regions}

\begin{figure*}
\includegraphics[width=1.0\textwidth, trim=10 60 40 90]{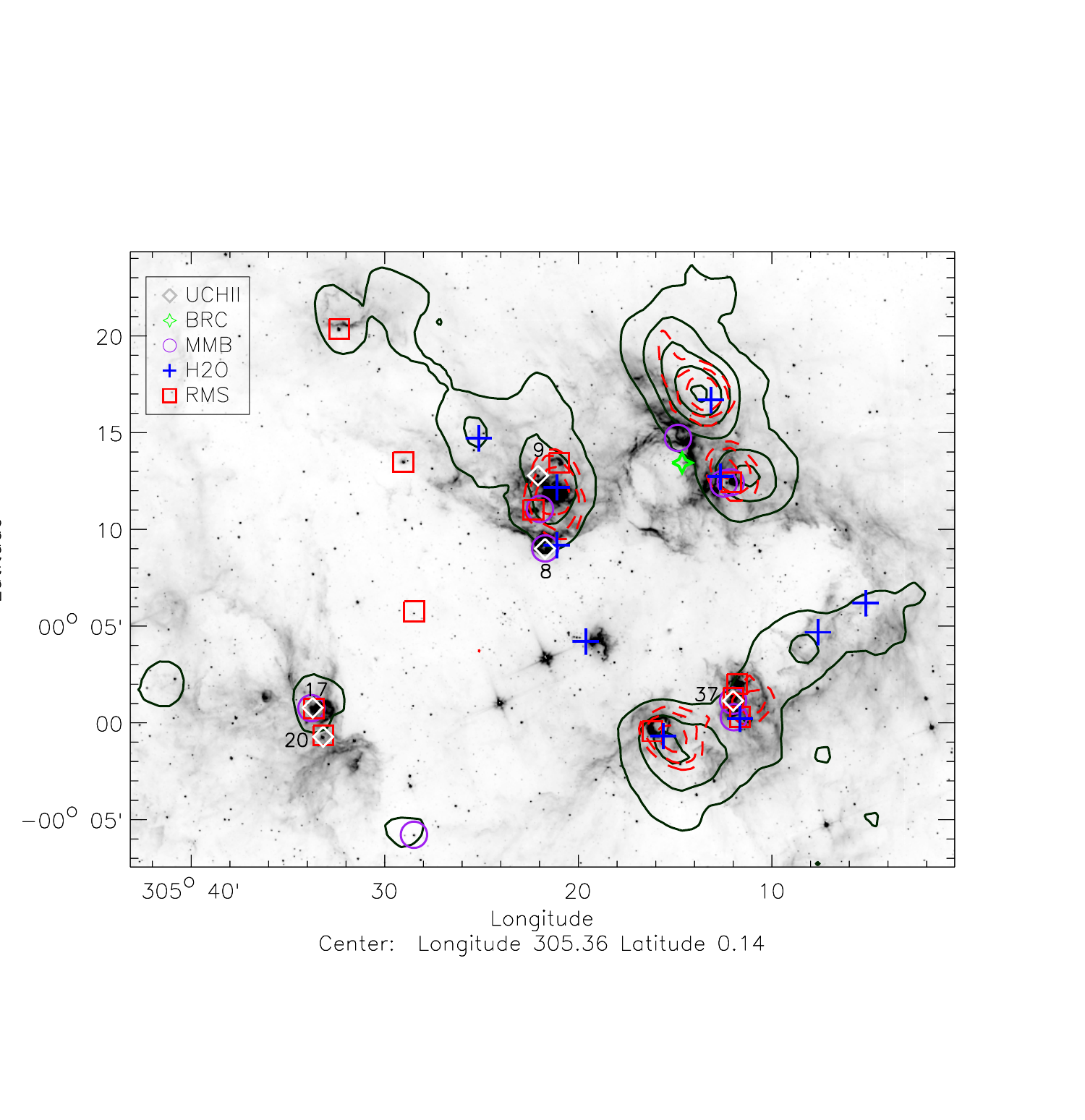} 
\caption{GLIMPSE 5.8\,$\umu$m background with the locations of \mbox{UC\,HII} regions (white diamonds), RMS survey sources (red squares), \water\ masers (blue crosses), methanol masers (purple circles) shown. Dense gas is traced by \NH\ (1,1) and (2,2) contours in black and red respectively starting at 3 and 9 and incrementing by 3 and 2 $\sigma$ respectively where $\sigma$ is the rms noise in the velocity integrated \NH\ maps ($\sim 0.01$\,K).}
\label{Multi}
\end{figure*}

We find that all five \mbox{UC\,HII} regions are located around the inner rim of the central cavity, within the PDR, towards areas where the mid-infrared emission is bright and compact (Fig.\,\ref{Multi}). We find no evidence of \mbox{UC\,HII} regions within the central cavity, outside the bounds of the PDR or towards the G305.2+0.2 region. Although all five \mbox{UC\,HII} regions are resolved we find four possess very simple morphology, primarily consisting of a very compact core which is associated with some low surface brightness diffuse emission and one source with an irregular distribution. The lack of structure seen toward the majority of these radio sources may suggest that they are still at a very early stage of their evolution.

\subsection{Comparison to \NH, masers and RMS data}

To explore the environment and high-mass star formation within G305 we compare the \mbox{UC\,HII} regions detected in this study to the dense molecular environment and star formation tracers taken from the literature.

Comparing the positions of the \mbox{UC\,HII} regions with respect to the dense molecular clouds located on the periphery of G305 we find four are located within the borders of the \NH\ (1,1) emission offset from the peak within the lower density region facing the central cavity. Source\,9 is an exception as it appears to be embedded within a dense \NH\ clump however, as mentioned in Paper\,1, this could be a projection effect and we may be viewing the cloud face on. We find that three sources (8, 9 and 37) are also on the borders of the higher excitation \NH\ (3,3) transition. Furthermore we find that the \NH\ clouds associated with \mbox{UC\,HII} regions have higher kinetic temperatures and line widths than clouds with no detectable \mbox{UC\,HII} region. This result is in agreement with studies such as \citealt{Urquhart2011B} where 80\% of massive star formation sites are associated \NH\ emission.
 
Water and methanol masers are both well correlated with star formation \citep{Urquhart2011B,Menten1991} and are thought to be associated with the very earliest stages (\ie\, prior to the formation of an HII region). We make use of the \water\ masers detected in Paper\,1 which have a positional accuracy of $<20\arcsec$ \citep{Walsh2011} and methanol masers from the MMB survey \citep{Green2009,Green2012,Caswell2009} which have a positional accuracy of 0.1\arcsec. 

We find nine 6.67\,GHz methanol masers within the field of view of G305 presented in Fig.\,4. Three of these methanol masers are associated with \mbox{UC\,HII} regions (8, 17 and 37) and are offset from the peak of the radio emission by $\lesssim 0.1$\,pc. This suggests they are associated with a nearby young stellar source rather than the \mbox{UC\,HII} region itself. It is unclear why two \mbox{UC\,HII} regions (9 and 20) have no methanol maser emission but it could indicate that they are older \citep{Ellingsen2007} and methanol masers have switched off. \mbox{UC\,HII} regions with no methanol maser emission are found to have sizes approximately twice that of those with methanol maser emission although caution should be taken as the diffuse emission has been removed in the data reduction. This also assumes source\,17 comprises more than one \mbox{UC\,HII} region, as suggested by the morphology, and indicates that the \mbox{UC\,HII} regions associated with methanol masers are younger than those that are not. Looking at the mid-infrared (Fig.\,3) it can been seen that the \mbox{UC\,HII} regions with methanol masers are associated with bright extended mid-infrared emission compared to \mbox{UC\,HII} regions without, which are found to be associated with isolated mid-infrared emission. There are six methanol masers that have no associated compact radio emission, which indicates that there is ongoing massive star formation in the region, embedded within the dense molecular gas, that have yet to create an observable \mbox{UC\,HII} region.

In Paper\,1, we present 16, 22\,GHz \water\ (6-5) masers (Table\,5 Paper\,1) towards G305. We find that these\water\ masers are anti-correlated with the position of \mbox{UC\,HII} regions. Comparison with the \NH\ data reveals that the \water\ masers are found predominantly embedded within the dense molecular clouds. This is consistent with the view that \water\ masers are tracing a younger evolutionary state than methanol masers and \mbox{UC\,HII} regions before the formation of a detectable HII region when the source is still deeply embedded as suggested by \cite{Garay1999}. 

Searching the Red MSX Source data base \citep{Urquhart2008}\footnote{http://www.ast.leeds.ac.uk/RMS/} we find 14 young massive stars (i.e.~HII regions or YSOs) that are positionally and kinematically associated with the G305 complex.\footnote{$^{13}$CO observations taken by the RMS survey team towards these sources show their velocities are well matched to the global velocity of G305 (\citealt{Urquhart2008B}.} The RMS survey is limited by the resolution and sensitivity of the MSX satellite which results in a detection limit of YSOs with a luminosity $>10^3$\,\lsun\ and mass of $> 7$\,\msun\ (see Fig.\,5 of \citealt{urquhart2011}). This sample of RMS sources highlights regions of recent massive star formation and these can be clearly seen to be associated with the PDR and the dense molecular gas in Fig.\,4. We find that RMS survey HII regions match the following \mbox{UC\,HII} sources detected here: 8, 17, 20 and 37 this further confirms their status as \mbox{UC\,HII} regions. Source\,9 has no RMS detection, this is likely due to its close proximity to a very bright-evolved HII region and the limited resolution of MSX. We find that all but one of the RMS surveys sources are found around the rim of G305 some distance from the central clusters Danks\,1 and 2. We are currently investigating the low-mass YSO component of G305 using \emph{Herschel} Hi-GAL survey (Faimali et al.~in prep) data and a YSO counting approach similar to \citet{Povich2011} study of Carina.

In a study carried out by \citet{Walsh1997, Walsh1998} the detection of five \mbox{UC\,HII} regions are reported towards G305 we find only one of these matches those detected here (Source\,17). This contradiction most likely arises because of a combination of the IRAS selection criteria used by Walsh et al.~ and the relatively short snapshot nature of their interferometry follow-up observations. With its limited resolution ($\sim 2\arcmin$) the IRAS selection has most likely mistaken more evolved compact HII regions, which appear unresolved, for \mbox{UC\,HII} regions. The snap shot nature of follow up radio observations \citep{Walsh1998} has then appeared to result in the peaks of evolved HII regions being mistakenly identified as \mbox{UC\,HII} regions. Work carried out by \citet{Longmore2007} found embedded infrared excess sources towards the G305.2+0.2 region. We find no compact radio emission associated with these young stellar sources or the G305.2+0.2 region. However, we do find evidence of interaction between the massive stars and molecular material in the region, as predicted by \citet{Longmore2007}, in the BRC (Source\,10) and also the enhancement of star formation indicators found along the boundary between ionised and neutral material of G305.2+0.2 (Fig.\,4).

\subsection{Star formation within G305}
It is clear that massive star formation is ongoing within G305 with the \mbox{UC\,HII} regions that we have identified providing evidence of at least five massive stars forming in the last few $10^5$\,yrs. In addition, RMS sources methanol and \water\ masers indicate many additional sites of massive star formation. These massive star formation tracers are primarily restricted to the periphery of the central cavity within the boundary of the dense molecular gas and PDR towards regions where the mid-infrared emission is bright and compact. 

Considering the distribution of stars and star formation across the G305 region, we can draw a broad-brush picture of the star formation history of G305. In the initial stages the central parts of the natal Giant Molecular Cloud collapsed to first form Danks\,2 ($\sim 3$\,Myr ago) and then Danks\,1 ($\sim 1.5$\,Myr ago; Paper\,2). For the last $\sim$3-6\,Myr the powerful UV radiation and winds from their population of massive stars has swept up the surrounding molecular material into the morphology we see in Fig.\,4. During this time a second generation of massive stars has formed, apparent by the presence of more evolved HII regions \citep{Clark2004}, around the cavity. Whether the massive stars responsible for these HII region were spontaneous or induced to form is uncertain.

It is around these HII regions that we find evidence of the most recent episodes of massive star formation traced by \mbox{UC\,HII} regions and methanol masers. \water\ maser emission towards the cores of \NH\ emission indicates that younger epochs of high and/or low mass star formation are taking place deeply embedded within the dense molecular material further still from the central cavity. Thus, it appears that the star formation in G305 is multi-generational and has at least partly occurred in spatially and temporally isolated bursts.

The morphology of the mid-infrared emission, dense molecular gas and the location of star formation tracers are suggestive of triggered star formation as proposed by \citet{Elmegreen1977}. Recent and ongoing star formation is enhanced and concentrated around the borders of four evolved HII regions (\citealt{Clark2004}; 305.551$-$0.005, 305.363+0.179, 305.254+0.204 and 305.202+0.002) between ionised and molecular gas. However, the morphology of a star-forming region is not considered sufficient evidence for triggered star formation since it may be that the radio or wind blown bubble is simply uncovering already existing star formation. In this regard our unbiased multi-wavelength datasets, encompassing the whole G305 complex, is particularly valuable in distinguishing between these possibilities via the identification of high and low mass YSOs.

If star formation were sequential, one would expect to find YSOs interior to the cavity but none outside, whereas they should be distributed throughout the GMC if this was not the case (cf. \citealt{Smith2010}). The star formation, including YSOs ($>7$\,\msun) identified by the RMS survey are found outside of the central cavity with exception to one source that is found at a projected distance of $>10$\,pc from the center of the complex. This phenomenon is also observed in Carina, where the YSOs are found to be concentrated towards the edges of bright rims \citep{Preibisch2011}.

We should note that there is one site of star formation observed towards the cavity centre, a \water\ maser that is associated with diffuse mid-infrared emission. However, this object is likely located on the far or near side of the cavity, rather than within the central cavity itself, as there is no evidence in the mid-IR morphology of any interaction with the winds from WR\,48A, Danks\,1 or Danks\,2. 

Further evidence of triggering is suggested by the interaction between the ionised and molecular material particularly in the G305.2+0.2 region where the BRC Source\,10 can be found. Whilst these observations are suggestive of triggering further study of the low-mass YSO component, as mentioned in Section\,5.2, and the interaction between the large scale radio emission associated with evolved HII regions and molecular material of G305 is essential for any determination of the global properties and to further explore the effect of massive star feedback in the region. 

\subsection{The current star formation rate in G305}

In Paper 2 we determined the mass and age of the two central stellar clusters Danks\,1 and 2, which are 8000$\pm$1500, 3000$\pm$800 M$_{\odot}$ and 1.5$^{+1.5}_{-0.5}$, 3$^{+3}_{-1}$ Myr respectively. If we make the assumption that the star formation within these clusters has proceeded constantly over their lifetimes then these masses and ages imply star formation rates of $\sim$ 10$^{-3}$--10$^{-2}$ M$_{\odot}$\,yr$^{-1}$, i.e.~at a similar rate to the Carina complex \citep{Povich2011}. In this section, we use our UC HII region detections to estimate the current star formation rate in G305 so that we may compare the current star formation rate to that of Danks\,1 and 2.

Before doing so we will briefly dwell upon the caveats involved in this calculation. Firstly, the stars powering the UC HII regions are by definition massive and thus we are forced to extrapolate the number of lower mass stars by integrating over the initial mass function (IMF). Secondly, the timescale for this star formation is not well known and will vary across the IMF from high mass to low mass. Low mass stars take $\sim$0.5 Myr to reach the pre-main sequence (PMS) phase (i.e.~the stage at which most of their stellar mass is assembled, \citealt{Evans2009,McKee2010,Offner2011}) whereas the higher mass stars that are the source of the ionising photons powering the UC HII regions reach the main sequence while still accreting. Both of these errors likely dominate over the error in determining the radio luminosity of the UC HIIs due to the uncertain distance of G305 (3.8$\pm$0.6 kpc, Paper 2).

However, obtaining even an order of magnitude estimate for the current star formation rate is instructive to compare with that determined for Danks\,1 and 2. Large differences in the star formation rate may indicate that the influence of Danks\,1 and 2 is either enhancing or suppressing star formation in the surrounding cloud. Our estimated star formation rate can also be compared with other star forming complexes, such as Carina, to place G305 in context with the rest of the Milky Way. 

To derive the current star formation rate over the past 0.5 Myr we assume that for each of our five \mbox{UC\,HII} regions, the most massive star present (shown in Table\,4) produces the majority of the ionising photon flux and is accompanied by a cluster of lower mass stars or YSOs. We then determine the total mass of stars and YSOs that may be present by extrapolating over a Salpeter IMF \citep{Salpeter1955} fixed to the mass of the most massive star in each UC HII region. We assume an upper and lower mass limit of50 and 0.1 \msun\ \citep{Robitaille2010}.

This extrapolation yields a total number of $\sim$1$ \times 10^{3}$ stars or YSOs which corresponds to a total stellar mass of $1.5 \times 10^{3}$\,M$_{\odot}$. This of the same order of magnitude as the mass in each of the Danks clusters and similar 
to the number of YSOs detected in a recent study of the Carina complex \citep{Preibisch2011}. As a consistency check we compared the number of the massive YSOs (i.e.~L$>10^{3}$ L$_{\odot}$) predicted by our IMF to the number of massive YSOs observed in the complex by the RMS and MMB surveys and found an agreement to within a factor of 2.

Assuming that the stars or YSOs have formed constantly over the last 0.5 Myr this implies an average star formation rate of $\sim$ 0.003 M$_{\odot}$\,yr$^{-1}$. We consider this rate to be a lower limit due to the likely incompleteness of our survey for UC HII regions located near to the bright and extended classical HII regions within G305, and also the fact that the $uv$ cut described in Sect.~\ref{sect:dr} means that we are insensitive to compact HII regions. We find that this estimate of the star formation rate in G305 is consistent with similar estimates made of the star formation rate in Carina \citep{Povich2011} and M17 \citep{Chomiuk2011}.

The current star formation rate that we estimate here is also comparable to that derived in Paper 2 for the Danks clusters. However, we do not consider it likely that G305 has sustained such a star formation rate over its history. If stars had been forming within the G305 complex at a constant rate of 5$\times$10$^{-3}$ M$_{\odot}$\,yr$^{-1}$ then over 5 Myr we would expect G305 to have formed $\sim$25\,000 M$_{\odot}$ of stars, which are clearly not detected. It is thus more likely that star formation within the G305 complex has proceeded in a punctuated fashion, with bursts of star formation at different epochs over the history of the complex. The morphology of G305 also suggests a multi-seeded star forming nature, with several distinct HII regions surrounding the main body of the complex that appear to be unrelated to the intense star formation happening at the rim of the central cavity. 

The average star formation rate for the entire Milky Way is estimated at $\sim$ 2 M$_{\odot}$\,yr$^{-1}$ \citep{Chomiuk2011,Davies2011}. With a star formation rates of between 10$^{-2}$--10$^{-3}$ M$_{\odot}$\,yr$^{-1}$ only a few tens to hundreds of complexes like G305, Carina and M17 could make up the bulk of the star formation in the Milky Way. The pinnacle is likely to be dominated by a few much larger complexes, for example the giant HII complexes identified by WMAP that make up half the total Galactic ionising flux \citep{Murray2010}. Just as the luminosity function of stars is dominated by the few most massive stars at the top of the IMF, Galactic star formation is likely dominated by a few massive star-forming complexes.

\section{Summary}

We have surveyed the star-forming complex G305 in search of compact 5.5 and 8.8\,GHz radio emission associated with \mbox{UC\,HII} regions. We find 71 compact radio sources randomly distributed across the observed field (Fig.\,\ref{CompactRadioGlimpse}). By matching to GLIMPSE mid-infrared data, we find 15 sources that are associated with G305, six of which we identify as candidate \mbox{UC\,HII} regions, one as a BRC, eight as stellar radio sources which leaves 56 background sources. 

Analysis of the six candidate \mbox{UC\,HII} regions reveals radio sources with physical properties consistent with known \mbox{UC\,HII} regions in five cases. We find typical \mbox{UC\,HII} properties for these sources, with source radii ranging from 0.04--0.1\,pc, emission measures $\sim\,\,2.56$--$10.3\times 10^{6}$\,pc cm$^{-6}$ and electron densities $\sim$\,0.34--$1.03\times 10^4$\,cm$^{-3}$.

By comparing to mid-infrared, \NH\ molecular line emission, \water\ and methanol masers and the RMS survey we comment on the star formation history of the region. We find that massive star formation over the last $10^5$\,yrs, traced by \mbox{UC\,HII} regions, has taken place around the periphery of the central cavity in close proximity to the ionised and molecular gas interface. Future star formation will most likely occur within the dense \NH\ clumps surrounding the central cavity. Massive star formation is ongoing and appears to have commenced with the formation of Danks 2 roughly 3-6\,Myr ago (Paper\,2) the effect of feedback from these clusters has yet to be determined. However, we do find morphological evidence suggestive of triggering. Using the detected \mbox{UC\,HII} regions we determine a lower limit for the SFR in G305 of $\sim$ 0.003\,M$_{\odot}$\,yr$^{-1}$. 

Future galactic plane surveys will soon provide similar or better quality multi-wavelength data across the extent of the Galaxy. These surveys include: HOPS \citep{Walsh2008}, CORNISH \citep{Purcell2010}, MALT90 \citep{Foster2011} and Hi-GAL \citep{Molinari2010A}. Using these large samples will allow rigorous statistical studies and resolve some of the long outstanding questions of massive star formation and evolution.

\section*{Acknowledgments}
We thank the anonymous referee for providing suggestions that greatly improved the clarity and quality of this paper. We would like to thank the Director and staff of the Paul Wild Observatory for their assistance with a pleasant and productive observing run. This research has made use of the NASA/ IPAC Infrared Science Archive (for access to GLIMPSE), which is operated by the Jet Propulsion Laboratory, California Institute of Technology, under contract with the National Aeronautics and Space Administration. This paper made use of information from the Red MSX Source survey database at www.ast.leeds.ac.uk/RMS that was constructed with support from the Science and Technology Facilities Council of the UK. We would


\label{lastpage}

\end{document}